\documentclass[showpacs,aps,graphicx,twocolumn]{revtex4}

\usepackage{graphicx}

\begin{document}

\title{Two-step hyperentanglement purification with the quantum-state-joining method\footnote{Published in Phys. Rev. A \textbf{90}, 052309 (2014)}}

\author{Bao-Cang Ren, Fang-Fang Du, and Fu-Guo Deng\footnote{Corresponding author: fgdeng@bnu.edu.cn} }

\address{ Department of Physics, Applied Optics Beijing Area Major Laboratory,
Beijing Normal University, Beijing 100875, China}

\date{\today }

\begin{abstract}

Hyperentanglement is a promising resource in quantum information
processing, especially for increasing the channel capacity of
long-distance quantum communication. Hyperentanglement purification
is an important method to obtain high-fidelity nonlocal
hyperentangled states from mixed hyperentangled states in a
long-distance quantum communication process with noisy channels.
Here, we present a two-step hyperentanglement purification protocol
for nonlocal mixed hyperentangled states with polarization bit-flip
errors and spatial-mode phase-flip errors, resorting to
polarization-spatial phase-check quantum nondemolition detectors and
the quantum-state-joining method (QSJM). With QSJM, the protocol can preserve
the states that are discarded in the previous hyperentanglement
purification protocols. It has the advantage of a high efficiency,
and it is useful for improving the entanglement of photon systems
with several degrees of freedom in long-distance high-capacity
quantum communication.

\end{abstract}

 \maketitle

\section{Introduction}

Entanglement is an essential quantum resource in quantum information
processing, and it improves the methods of manipulating and
transmitting information in quantum computation and quantum
communication \cite{QC}. The maximally entangled photon system, used
as a quantum channel, is a core resource in  quantum communication
\cite{QT1,DC1,DC2,QKD,QKD2,QKD3,QKD4,QSDC1,QSDC2,QSS1,QSS2,QSS3}.
With maximally entangled photon pairs, the two remote parties in
quantum communication, say Alice and Bob, can implement faithful
teleportation
 \cite{QT1}. That is, they teleport the unknown quantum state of a single
particle without moving the particle itself \cite{QT1}. Also, Alice
can transmit two or more bits of information to Bob by moving only a
photon back and forth with quantum dense coding protocols based on a
maximally entangled photon pair \cite{DC1,DC2}. They can create a
private key with maximally entangled photon pairs by resorting to
quantum key distribution protocols \cite{QKD,QKD2,QKD3}, even in the
case with a collective-noise optical-fiber channel \cite{QKD4}. With
quantum secure direct communication protocols based on entangled
photon pairs \cite{QSDC1,QSDC2}, Alice and Bob can also transmit their
secret message directly, without creating a private key in
advance. With entanglement-based quantum secret-sharing protocols
\cite{QSS1,QSS2,QSS3}, the parties in quantum communication can
share some private keys with a potentially dishonest party.

Hyperentanglement, defined as the entanglement in several degrees of
freedom (DOFs) of a quantum system \cite{heper1,heper2,heper3}, has
attracted much attention for improving the power of quantum
computation (e.g., hyperparallel photonic quantum computation
\cite{hypercnot,hypercnot1}) and increasing the channel capacity of
long-distance quantum communication, such as photonic superdense
coding with polarization-orbital-angular-momentum
hyperentanglement \cite{HESC}, quantum teleportation with photon
pairs entangled in two DOFs \cite{kerr}, entanglement swapping based
on photonic spatial-polarization hyperentanglement
\cite{kerr,HBSA2}, hyperentangled Bell-state analysis
\cite{kerr,HBSA2,HBSA,HBSA1,HBSA3}, hyperentanglement concentration
\cite{hyperecp,hyperecp2}, and so on. With two-photon
hyperentanglement, the Greenberger-Horne-Zeilinger (GHZ) theorem has
been generalized to the case with only two entangled particles
\cite{GHZ}, and the first two-particle all-or-nothing test of
local realism has been demonstrated experimentally with a linear
optics setup, without resorting to a noncontextuality assumption
\cite{AVN}. Hyperentanglement has also been used to assist
long-distance quantum communication in one DOF of photon systems,
such as a quantum repeater \cite{wangrepeater}, complete Bell-state
analysis \cite{BSA1kwiat,BSA2walborn,BSA3,BSA4,EPPsheng2}, and
deterministic entanglement purification
\cite{EPPsheng2,EPPsheng3,EPPlixh,EPPdeng,DEPPsheng14} on the
polarizations of photon pairs. In long-distance quantum
communication protocols, the photon signals can be transmitted
no more than several hundreds of kilometers over an optical fiber or
a free space, and quantum repeaters are required in this condition.
The (hyper)entangled photon systems are always produced locally,
and they  inevitably suffer from environment noise in their
distribution process and storage process in quantum communication.
In general, the maximally (hyper)entangled photon states may decay
into less entangled pure states or even into mixed states, which
decreases the fidelity and the security of long-distance quantum
communication protocols.

Entanglement purification is an interesting passive way to suppress
the effect of environmental noise in quantum communication
processes. It is used to distill some quantum systems in high-fidelity
entangled states from those in less-entangled mixed states
\cite{EPPsheng2,EPPsheng3,EPPlixh,EPPdeng,DEPPsheng14,EPP1,EPP2,EPP3,EPPexperiment,EPPsimon,EPPsheng1,EPPdengpra,EPPsheng13,MEPP,MEPP2,HEPP,EPPcaowang,EPPshengzhou,eppa1,eppa2,eppa3}.
In 1996, Bennett \emph{et al.} \cite{EPP1} introduced the first
entanglement purification protocol (EPP) for quantum systems in a
Werner state with quantum controlled-not (CNOT) gates.  In 2001, Pan
\emph{et al}. \cite{EPP3} presented a polarization EPP with linear
optics. In 2002, Simon and Pan proposed a polarization EPP assisted
by hyperentanglement with the available parametric down-conversion
(PDC) resource \cite{EPPsimon}, which was demonstrated by Pan
\emph{et al.}  \cite{EPPexperiment} in 2003. In 2008, Sheng \emph{et
al}. \cite{EPPsheng1} proposed an efficient polarization EPP with
hyperentanglement based on a PDC resource, resorting to cross-Kerr
nonlinearity.  In 2010, Sheng and Deng \cite{EPPsheng2}
introduced the concept of deterministic entanglement purification
for two-photon systems with hyperentanglement. Subsequently, some
interesting deterministic EPPs were proposed
\cite{EPPsheng3,EPPlixh,EPPdeng,DEPPsheng14} with linear optical
elements.  The EPPs for multipartite entangled-photon systems in
mixed-polarization GHZ states have also been investigated with
nonlinear optical elements in the past few years \cite{MEPP,MEPP2}.
In 2013, a hyperentanglement purification protocol (hyper-EPP) was
proposed for two-photon systems in mixed polarization-spatial
hyperentangled Bell states with polarization bit-flip errors and
spatial-mode bit-flip errors resorting to nonlinear optical elements
\cite{HEPP}. However, it is implemented by purifying the
polarization states and the spatial-mode states of photon pairs
independently. In 2013, Vitelli \emph{et al}. \cite{QSJ}
demonstrated experimentally the quantum-state-joining process with
linear optical elements, which can combine the two-dimensional
quantum states of two input photons into an output single photon.

In this paper, we present a two-step high-efficiency hyper-EPP for
nonlocal photon systems in  mixed polarization-spatial
hyperentangled states, resorting to the
 polarization-spatial phase-check quantum nondemolition detectors
(P-S-QNDs) and the quantum-state-joining method (QSJM). Our P-S-QND
is implemented with the hybrid CNOT gate \cite{hypercnot1} based on
the giant optical circular birefringence (GOCB) of the quantum-dot
(QD) spin inside a double-sided optical microcavity (a double-sided
QD-cavity system). Our QSJM is also constructed with the GOCB of
a double-sided QD-cavity system. With our QSJM, mixed
hyperentangled states with only one DOF in the preserving condition
can be combined into mixed hyperentangled states with two
DOFs in the preserving condition, and the efficiency of our two-step hyper-EPP can be greatly improved
by preserving the states that are discarded in the previous
hyper-EPP without QSJM. QSJM is very useful for improving the
entanglement of photon systems in long-distance quantum communication.

This paper is organized as follows: In Sec. \ref{sec21}, we
introduce the  GOCB of a quantum-dot  spin inside a double-sided
optical microcavity. Based on GOCB, we give the details of our QSJM
and our P-S-QND in Secs. \ref{sec22} and  \ref{sec23}, respectively.
In Sec. \ref{sec3}, we introduce our two-step high-efficiency
hyper-EPP for nonlocal two-photon systems with QSJM. Its
generalization  for hyperentangled three-photon GHZ states is given
in Sec. \ref{sec4}. A discussion and a summary are given in Sec.
\ref{sec5}. In Appendixes \ref{appendixa} and \ref{appendixb}, we
give the QSJM for hyperentangled Bell states and a detailed
calculation for the fidelities and the efficiencies of P-S-QND and
QSJM, respectively.

\section{Quantum-state-joining method and polarization-spatial phase-check QND}
\label{sec2}

\subsection{Giant optical circular birefringence of a quantum-dot electron spin in a double-sided optical resonant
microcavity} \label{sec21}

\begin{figure}[htbp]             
\centering\includegraphics[width=7.22 cm]{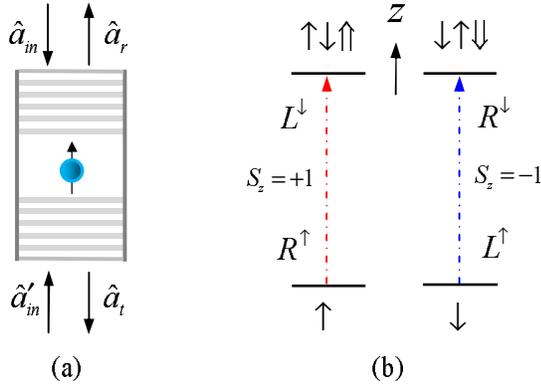} \caption{(Color
online) The optical transition of a QD embedded in a double-sided
microcavity with circularly  polarized  photons. (a) A double-sided
QD-cavity system. The QD is located at the center of a double-sided
cavity for maximal light-matter coupling. (b) The spin-dependent
optical transitions of a negatively charged exciton $X^-$ with
circularly polarized photons. $L^\uparrow$ ($L^\downarrow$) and
$R^\uparrow$ ($R^\downarrow$) represent the left- and right-circularly
polarized lights with their input directions parallel
(antiparallel) to the z direction, respectively.
$\uparrow\downarrow\Uparrow$ ($\downarrow\uparrow\Downarrow$)
represents the spin state $|+\frac{3}{2}\rangle$
($|-\frac{3}{2}\rangle$) of the negatively charged exciton $X^-$.
$\uparrow$ ($\downarrow$) represents the spin state
$|+\frac{1}{2}\rangle$ ($|-\frac{1}{2}\rangle$) of the excess
electron in a QD.} \label{figure_1}
\end{figure}

The QSJM and the P-S-QND in our proposal are both constructed with
the optical property of a singly charged QD [e.g., a self-assembled
In(Ga)As QD or GaAs interfacial QD] embedded in a double-sided
optical resonant microcavity, as shown in Fig. \ref{figure_1}(a).
The two distributed Bragg reflectors of the double-sided optical
resonant microcavity are partially reflective and low loss for
on-resonance transmission, which supports both polarization
modes of photon systems \cite{cavity}. If an excess electron is
injected into the QD, the singly charged QD has optical
resonance with the creation of a negatively charged exciton $X^-$,
which consists of two antiparallel electron spins bound to one hole
\cite{QD}, and it shows spin-dependent transitions with circularly
polarized lights according to Pauli's exclusion principle \cite{QD1}
[shown in Fig. \ref{figure_1}(b)]. If the excess electron is in the
spin state $|\uparrow\rangle$, a circularly polarized photon with
spin $S_z=+1$ is absorbed to create the negatively charged
exciton $X^-$ in the state $|\uparrow\downarrow\Uparrow\rangle$.
However, for the excess electron-spin state $|\downarrow\rangle$, a
circularly polarized photon with spin $S_z=-1$ is absorbed to
create the negatively charged exciton $X^-$ in the state
$|\downarrow\uparrow\Downarrow\rangle$. Here $|\Uparrow\rangle$
($|\Downarrow\rangle$) represents the heavy-hole spin
$|+\frac{3}{2}\rangle$ ($|-\frac{3}{2}\rangle$), and
$|\uparrow\rangle$ ($|\downarrow\rangle$) represents the excess
electron spin $|+\frac{1}{2}\rangle$ ($|-\frac{1}{2}\rangle$).

The input-output optical property of the double-sided QD-cavity
system can be described by the Heisenberg equations of motion for
the cavity field operator $\hat{a}$ and the $X^-$ dipole operator
$\hat{\sigma}_-$ in the interaction picture \cite{QD2,QD3},
\begin{eqnarray}                           
\frac{d\hat{a}}{dt}&=&-[i(\omega_c-\omega)+\kappa+\frac{\kappa_s}{2}]\hat{a}-g\hat{\sigma}_-\nonumber\\
&& -\sqrt{\kappa}\,\hat{a}_{in}-\sqrt{\kappa}\,\hat{a}'_{in}, \nonumber\\
\frac{d\hat{\sigma}_-}{dt}&=&-[i(\omega_{X^-}-\omega)+\frac{\gamma}{2}]\hat{\sigma}_--g\hat{\sigma}_z\hat{a},\nonumber\\
\hat{a}_{r}&=&\hat{a}_{in}+\sqrt{\kappa}\,\hat{a},\nonumber\\
\hat{a}_{t}&=&\hat{a}'_{in}+\sqrt{\kappa}\,\hat{a},
\end{eqnarray}
where $\omega_c$, $\omega$, and $\omega_{X^-}$ are the frequencies
of the cavity mode, the input photon, and the $X^-$ transition,
respectively. $g$ is the coupling strength of the negatively charged
exciton $X^-$ and the cavity mode. $\gamma/2$ is the decay rate of
$X^-$. $\kappa$ is the decay rate of the cavity field mode, and
$\kappa_s/2$ is the decay rate of the cavity field mode to the
cavity side leakage mode. $\hat{a}_{in}$, $\hat{a}'_{in}$ and
$\hat{a}_{r}$, $\hat{a}_{t}$ are the input and output field
operators of the double-sided QD-cavity system, respectively.

In the weak excitation condition, $X^-$ is dominantly in the ground
state with $\langle\sigma_z\rangle=-1$, and the reflection
[$r(\omega)$] and transmission [$t(\omega)$] coefficients of the
double-sided QD-cavity system are described by \cite{QD3}
\begin{eqnarray}                           
r(\omega)&\!=\!&1+t(\omega),\nonumber\\
t(\omega)&\!=\!&\frac{-\kappa[i(\omega_{X^-}-\omega)+\frac{\gamma}{2}]}{[i(\omega_{X^-}-\omega)
+\frac{\gamma}{2}][i(\omega_c-\omega)+\kappa+\frac{\kappa_s}{2}]+g^2}.\;\;\;\;\;\;
\end{eqnarray}
In the resonant condition ($\omega_c=\omega_{X^-}=\omega_0$), if the
coupling strength is $g=0$ (the QD is decoupled from the cavity), the
reflection and transmission coefficients can be expressed as
\cite{QD3}
\begin{eqnarray}                           
r_0(\omega)&=&\frac{i(\omega_0-\omega)+\frac{\kappa_s}{2}}{i(\omega_0-\omega)+\kappa+\frac{\kappa_s}{2}},\nonumber\\
t_0(\omega)&=&\frac{-\kappa}{i(\omega_0-\omega)+\kappa+\frac{\kappa_s}{2}}.
\end{eqnarray}
In the strong-coupling regime [$g>(\kappa,\gamma)$] and the resonant
condition ($\omega_c=\omega_{X^-}=\omega_0\approx\omega$), if the
cavity side leakage is neglected, the reflection and transmission
coefficients can achieve $|r|\rightarrow1$, $|r_0|\rightarrow0$ and
$|t|\rightarrow0$, $|t_0|\rightarrow1$.

As the photonic circular polarization   usually depends on the
direction of propagation, the photon in state
$|R^\uparrow\rangle$ or $|L^\downarrow\rangle$ has spin
$S_z=+1$, and the photon in state $|R^\downarrow\rangle$ or
$|L^\uparrow\rangle$ has spin $S_z=-1$. Here $L^\uparrow$
($L^\downarrow$) and $R^\uparrow$ ($R^\downarrow$) represent the
left- and right-circularly polarized lights with their input
directions parallel (antiparallel) to the z direction, respectively,
shown in Fig. \ref{figure_1}(b). For the electron spin state
$|\uparrow\rangle$, the input circularly polarized photon
$|R^\uparrow\rangle$ ($|L^\downarrow\rangle$) is reflected as
$|L^\downarrow\rangle$ ($|R^\uparrow\rangle$) by the QD-cavity
system with the QD coupled to the cavity, while the input circularly
polarized photon $|R^\downarrow\rangle$ ($|L^\uparrow\rangle$) is
transmitted through the  QD-cavity system with a phase shift
relative to the reflected photon (QD is decoupled from the cavity).
For the electron spin state $|\downarrow\rangle$, the input
circularly polarized photon $|R^\uparrow\rangle$
($|L^\downarrow\rangle$) is transmitted through the QD-cavity
system, and the input circularly polarized photon
$|R^\downarrow\rangle$ ($|L^\uparrow\rangle$) is reflected by the
QD-cavity system. That is, the reflection and transmission rules
of the photon polarization states can be summarized as follows
\cite{hypercnot1,QD3}:
\begin{eqnarray}                           
&&\!\!\!\!|R^\uparrow, i_2, \uparrow\rangle \rightarrow
|L^\downarrow, i_2, \uparrow\rangle,\;\;\;\;\;\,
|L^\downarrow, i_1, \uparrow\rangle \rightarrow |R^\uparrow, i_1, \uparrow\rangle,\nonumber\\
&&\!\!\!\!|R^\uparrow, i_2, \downarrow\rangle \rightarrow
-|R^\uparrow, i_1\downarrow\rangle,\;\;\; |L^\downarrow,
i_1,\downarrow\rangle \rightarrow -|L^\downarrow, i_2,
\downarrow\rangle,\nonumber\\
 &&\!\!\!\!|R^\downarrow, i_1,
\uparrow\rangle \rightarrow -|R^\downarrow, i_2,
\uparrow\rangle,\;\;\,
|L^\uparrow, i_2, \uparrow\rangle \rightarrow -|L^\uparrow, i_1, \uparrow\rangle,\nonumber\\
&&\!\!\!\!|R^\downarrow, i_1, \downarrow\rangle \rightarrow
|L^\uparrow, i_1, \downarrow\rangle,\;\;\;\;\;\; |L^\uparrow, i_2,
\downarrow\rangle \rightarrow |R^\downarrow, i_2,
\downarrow\rangle.\;\;\;\;\;\;\;\;\label{eq4}
\end{eqnarray}
Here $i_1$ and $i_2$ ($i=a, b$) are the two spatial modes of
photon $i$ (shown in Fig. \ref{figure2}).

\subsection{Quantum-state-joining method} \label{sec22}

\begin{figure}[!h] 
\centering
\includegraphics[width=8 cm]{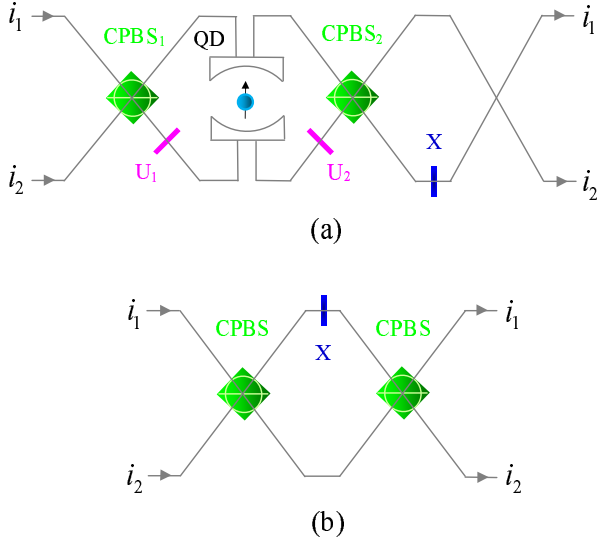}
\caption{(Color online) (a) Schematic diagram of our
quantum-state-joining method (QSJM). (b) Schematic diagram of the
swap gate between the spatial-mode state and the polarization state
of a photon.  QD represents a double-sided QD-cavity system.
CPBS$_i$ ($i=1,2$) represents a polarizing beam splitter in the
circular basis, which transmits the right-circular polarization
photon $|R\rangle$ and reflects the left-circular polarization
photon $|L\rangle$. U$_i$ ($i=1,2$) represents a wave
plate which is used to perform a polarization phase-flip operation
$U=-|R\rangle\langle R|-|L\rangle\langle L|$ on a photon. X
represents a half-wave plate which is used to perform a polarization
bit-flip operation $\sigma_x^P=|R\rangle\langle L|+|L\rangle\langle
R|$ on the photon passing through it. $i_1$ and $i_2$ represent the
two spatial modes of photon $i$ ($i=a,b$).} \label{figure2}
\end{figure}

The QSJM for photonic states introduced here is used to transfer the
polarization state of photon $A$ to the polarization state of photon
$B$ without disturbing the spatial-mode state of photon $B$,
resorting to the reflection-transmission optical property of the
double-sided QD-cavity system, as shown in Fig. \ref{figure2}(a).
The initial states of the two photons, $A$ and $B$, are
\begin{eqnarray}                      
|\phi_A\rangle &=& (\alpha_1|R\rangle+\alpha_2|L\rangle)_A(\gamma_1|a_1\rangle+\gamma_2|a_2\rangle),\nonumber\\
|\phi_B\rangle &=&
(\beta_1|R\rangle+\beta_2|L\rangle)_B(\delta_1|b_1\rangle+\delta_2|b_2\rangle).
\end{eqnarray}
The excess electron spin in the QD is prepared in the state
$|+\rangle_e=\frac{1}{\sqrt{2}}(|\uparrow\rangle+|\downarrow\rangle)_e$.

We put photon $A$ into the circularly polarizing beam splitter
CPBS$_1$, wave plate U$_1$, QD, U$_2$, CPBS$_2$, and
half-wave plate X in sequence, as shown in Fig. \ref{figure2}(a).
After photon $A$ passes through the quantum circuit shown in
Fig. \ref{figure2}(a), the state of the quantum system composed of
a QD and photon $A$ is transformed from $|\phi_{Ae}\rangle_0$ to
$|\phi_{Ae}\rangle_1$. Here
\begin{eqnarray}                           
|\phi_{Ae}\rangle_0&=&|+\rangle_e\otimes|\phi_A\rangle,\nonumber\\
|\phi_{Ae}\rangle_1&=&\frac{1}{\sqrt{2}}\big[|R\rangle_A(\alpha_1|\uparrow\rangle+\alpha_2|\downarrow\rangle)_e
+|L\rangle_A(\alpha_2|\uparrow\rangle\;\;\;\;\;\;\;\;\nonumber\\
&&+\alpha_1|\downarrow\rangle)_e\big](\gamma_1|a_1\rangle+\gamma_2|a_2\rangle).\label{eq5}
\end{eqnarray}
By measuring the polarization state of photon $A$ in the
orthogonal basis $\{|R\rangle,|L\rangle\}$, the polarization state
of photon $A$ is transferred to the excess electron spin state
in the QD.

We assume the excess electron spin in the QD is in the state
$|\phi\rangle_e=(\alpha_1|\uparrow\rangle+\alpha_2|\downarrow\rangle)_e$
(i.e., the polarization state of photon $A$ is projected into
state $|R\rangle$). Photon $B$ is put into the quantum
circuit shown in Fig. \ref{figure2}(a) after a Hadamard operation $[\vert \uparrow\rangle \rightarrow \frac{1}{\sqrt{2}}(\vert\uparrow\rangle + \vert \downarrow\rangle)$, $\vert \downarrow\rangle \rightarrow \frac{1}{\sqrt{2}}(\vert\uparrow\rangle - \vert \downarrow\rangle)]$
is performed on the excess electron spin in the QD, and the state of the
quantum system composed of the QD and photon $B$ is transformed from
$|\phi_{Be}\rangle_1$ to $|\phi_{Be}\rangle_2$. Here
\begin{eqnarray}                           
|\phi_{Be}\rangle_1&=&|\phi_B\rangle\otimes|\phi_e\rangle,\nonumber\\
|\phi_{Be}\rangle_2&=&\big[\alpha'_1|\uparrow\rangle_e(\beta_1|R\rangle+\beta_2|L\rangle)_B+\alpha'_2|\downarrow\rangle_e\nonumber\\
&&(\beta_2|R\rangle+\beta_1|L\rangle)_B\big](\delta_1|b_1\rangle+\delta_2|b_2\rangle),\;\;\;\;\;\;\;\;
\end{eqnarray}
where $\alpha'_1=\frac{1}{\sqrt{2}}(\alpha_1+\alpha_2)$ and
$\alpha'_2=\frac{1}{\sqrt{2}}(\alpha_1-\alpha_2)$. After the
Hadamard operations are  performed on the polarization DOF of
photon $B$ and the excess electron spin $e$ in sequence, we have
photon $B$ pass through the quantum circuit shown in Fig.2(a)
again, and the state of the quantum system $Be$ is changed to
$|\phi_{Be}\rangle_3$. Here
\begin{eqnarray}                          
|\phi_{Be}\rangle_3&=&\big[\alpha_1|R\rangle_B(\beta'_1|\uparrow\rangle+\beta'_2|\downarrow\rangle)_e+\alpha_2|L\rangle_B\nonumber\\
&&(\beta'_2|\uparrow\rangle+\beta'_1|\downarrow\rangle)_e\big](\delta_1|b_1\rangle+\delta_2|b_2\rangle),
\end{eqnarray}
where $\beta'_1=\frac{1}{\sqrt{2}}(\beta_1+\beta_2)$ and
$\beta'_2=\frac{1}{\sqrt{2}}(\beta_1-\beta_2)$. Finally, a Hadamard
operation is performed on the excess electron spin $e$, and the
state of the quantum system $Be$ is changed from $|\phi_{Be}\rangle_3$ to be
$|\phi_{Be}\rangle_4$. Here
\begin{eqnarray}                           
|\phi_{Be}\rangle_4&=&\big[\beta_1|\uparrow\rangle_e(\alpha_1|R\rangle+\alpha_2|L\rangle)_B+\beta_2|\downarrow\rangle_e\nonumber\\
&&(\alpha_1|R\rangle-\alpha_2|L\rangle)_B\big](\delta_1|b_1\rangle+\delta_2|b_2\rangle).
\end{eqnarray}
By measuring the excess electron spin $e$ in the orthogonal basis
$\{|\!\!\uparrow\rangle,|\!\!\downarrow\rangle\}$, the state of the
excess electron spin $e$
[$|\phi\rangle_e=(\alpha_1|\uparrow\rangle+\alpha_2|\downarrow\rangle)_e$]
is transferred  to the polarization state of photon $B$ without
disturbing its spatial-mode state. If the polarization DOF of photon
$A$ is projected into state $|L\rangle$, a bit-flip operation
($\sigma_x^P=|R\rangle\langle L|+|L\rangle\langle R|$) is performed
on the polarization DOF of photon $B$, and a phase-flip operation
($\sigma^P_z=|R\rangle\langle R| - |L\rangle\langle L|$) is
performed on the polarization DOF of photon $B$ if the excess
electron spin $e$ is projected into state
$|\!\!\downarrow\rangle_e$. After the conditional operations are
performed on the polarization DOF of photon $B$, the state of photon
$B$ is changed to
$|\phi_B\rangle_f=(\alpha_1|R\rangle+\alpha_2|L\rangle)_B(\delta_1|b_1\rangle+\delta_2|b_2\rangle)$.
This is the result of the QSJM for combining the polarization state
of photon $A$ and the spatial-mode state of photon $B$ into an
output single photon.

The QSJM can also be used to transfer the spatial-mode state of
photon $A$ to the polarization state of photon $B$. This requires
that photon $A$ passes through the quantum circuit shown in Fig.
\ref{figure2}(b), which can swap the polarization state and the
spatial-mode state of photon $A$. Subsequently, we can transfer the
polarization state of photon $A$ to the polarization state of photon
$B$ with the quantum circuit shown in Fig. \ref{figure2}(a). Then
the result of the QSJM for combining the spatial-mode state of
photon $A$ and the spatial-mode state of photon $B$ into a single
output photon is achieved.

\subsection{Polarization-spatial phase-check QND} \label{sec23}

\begin{figure}[!h]
\centering
\includegraphics[width=8 cm,angle=0]{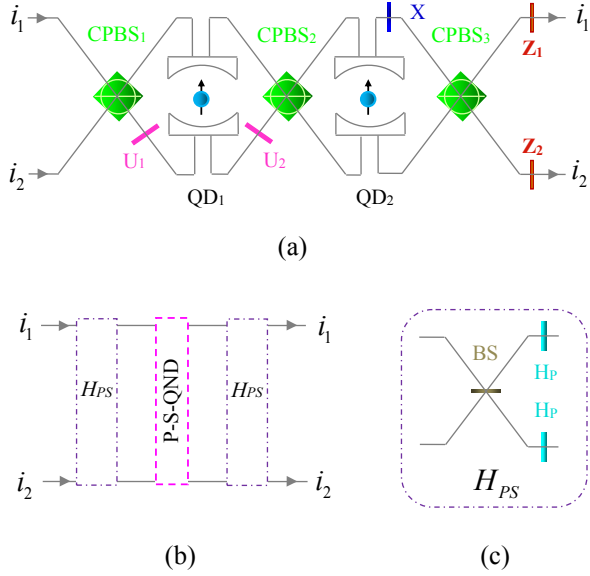}
\caption{(Color online) (a) Schematic diagram of the
polarization-spatial phase-check QND (P-S-QND). (b) Schematic
diagram of the polarization-spatial parity-check QND. (c) Schematic
diagram of $H_{PS}$, which is used to perform the Hadamard operation on both the polarization and the spatial-mode DOFs of a photon.  BS represents a 50:50 beam splitter which is
used to perform the Hadamard operation $[\vert i_1\rangle\rightarrow \frac{1}{\sqrt{2}}(\vert i_1\rangle + \vert i_2\rangle)$, $\vert i_2\rangle\rightarrow \frac{1}{\sqrt{2}}(\vert i_1\rangle - \vert i_2\rangle)]$ on the spatial-mode DOF of a
photon. Z$_i$ ($i=1,2$) represents a half-wave plate which is used
to perform the polarization phase-flip operation
$\sigma_z^P=-|R\rangle\langle R|+|L\rangle\langle L|$ on a photon.
$H_P$ represents a half-wave plate which is used to perform the
Hadamard operation $[\vert R\rangle\rightarrow \frac{1}{\sqrt{2}}(\vert R\rangle + \vert L\rangle)$, $\vert L\rangle\rightarrow \frac{1}{\sqrt{2}}(\vert R\rangle - \vert L\rangle)]$ on the polarization DOF of a photon.}
\label{figure3}
\end{figure}

Now, we introduce the construction of the polarization-spatial
phase-check QND, which is used to distinguish the
hyperentangled Bell states with the relative phase $0$ from those
with the relative phase $\pi$ in both the polarization and the
spatial-mode DOFs. Our P-S-QND is implemented with the hybrid CNOT
gate introduced in our previous work \cite{hypercnot1}, and its setup is shown in
Fig. \ref{figure3}(a). The 16 polarization-spatial hyperentangled
Bell states are defined as
$|\phi_{kl}\rangle_{AB}=|\phi_k\rangle_{AB}^P\otimes|\phi_l\rangle_{AB}^S$
($k,l=1,2,3,4$), where
\begin{eqnarray}                           
|\phi_1\rangle_{AB}^P&=&\frac{1}{\sqrt{2}}(|RR\rangle+|LL\rangle)_{AB},\nonumber\\
|\phi_2\rangle_{AB}^P&=&\frac{1}{\sqrt{2}}(|RR\rangle-|LL\rangle)_{AB},\nonumber\\
|\phi_3\rangle_{AB}^P&=&\frac{1}{\sqrt{2}}(|RL\rangle+|LR\rangle)_{AB},\nonumber\\
|\phi_4\rangle_{AB}^P&=&\frac{1}{\sqrt{2}}(|RL\rangle-|LR\rangle)_{AB},\nonumber\\
|\phi_1\rangle_{AB}^S&=&\frac{1}{\sqrt{2}}(|a_1b_1\rangle+|a_2b_2\rangle),\nonumber\\
|\phi_2\rangle_{AB}^S&=&\frac{1}{\sqrt{2}}(|a_1b_1\rangle-|a_2b_2\rangle),\nonumber\\
|\phi_3\rangle_{AB}^S&=&\frac{1}{\sqrt{2}}(|a_1b_2\rangle+|a_2b_1\rangle),\nonumber\\
|\phi_4\rangle_{AB}^S&=&\frac{1}{\sqrt{2}}(|a_1b_2\rangle-|a_2b_1\rangle).
\end{eqnarray}
Here the subscript $AB$ represents the photon pair $AB$, and the
superscripts $P$ and $S$ represent the polarization and the
spatial-mode DOFs of a photon system, respectively.

The initial states of the excess electron spins $e_1$ in QD$_1$ and
$e_2$ in QD$_2$ are
$|+\rangle_{e_1}=\frac{1}{\sqrt{2}}(|\uparrow\rangle+|\downarrow\rangle)_{e_1}$
and
$|+\rangle_{e_2}=\frac{1}{\sqrt{2}}(|\uparrow\rangle+|\downarrow\rangle)_{e_2}$,
respectively. If we have photons $A$ and $B$ pass through the
quantum circuit of the hybrid CNOT gate shown in Fig. \ref{figure3}(a)
in sequence,  the state of the quantum system composed of photon pair
$AB$ and the excess electron spins $e_1$ and $e_2$ (in QD$_1$ and QD$_2$,
respectively) is transformed into
\begin{eqnarray}                           
|\phi_{k_1l_1}\rangle_{AB}|+\rangle_{e_1}|+\rangle_{e_2}&\rightarrow&|\phi_{k_1l_1}\rangle_{AB}|+\rangle_{e_1}|-\rangle_{e_2},\nonumber\\
|\phi_{k_1l_2}\rangle_{AB}|+\rangle_{e_1}|+\rangle_{e_2}&\rightarrow&|\phi_{k_1l_2}\rangle_{AB}|+\rangle_{e_1}|+\rangle_{e_2},\nonumber\\
|\phi_{k_2l_1}\rangle_{AB}|+\rangle_{e_1}|+\rangle_{e_2}&\rightarrow&|\phi_{k_2l_1}\rangle_{AB}|-\rangle_{e_1}|-\rangle_{e_2},\nonumber\\
|\phi_{k_2l_2}\rangle_{AB}|+\rangle_{e_1}|+\rangle_{e_2}&\rightarrow&|\phi_{k_2l_2}\rangle_{AB}|-\rangle_{e_1}|+\rangle_{e_2},\;\;\;\;\;\;\;\;
\end{eqnarray}
where $k_1(l_1)=1,3$,  $k_2(l_2)=2,4$, and
$|\pm\rangle=\frac{1}{\sqrt{2}}(|\!\!\uparrow\rangle\pm|\!\!\downarrow\rangle)$.
By measuring the spin states of excess electrons $e_1$ and $e_2$
in the orthogonal basis $\{|+\rangle, |-\rangle\}$
\cite{hypercnot1}, one can distinguish the hyperentangled Bell states
with the relative phase $0$ from those with the relative phase
$\pi$ in both the polarization and the spatial-mode DOFs. If the
excess electron spin $e_1$ ($e_2$) is projected into state
$|+\rangle_{e_1}$ ($|-\rangle_{e_2}$), the polarization
(spatial-mode) state of the hyperentangled Bell state has the
relative phase $0$.  Otherwise the polarization (spatial-mode)
state of the hyperentangled Bell state has the relative phase
$\pi$.

If we perform Hadamard operations on both the spatial-mode and the
polarization DOFs of photons $A$ and $B$ before (and after) we
put them into the quantum circuit of our P-S-QND  shown in Fig.
\ref{figure3}(b), the odd-parity mode (i.e., $\vert \phi_3\rangle^P_{AB}$, $\vert \phi_4\rangle^P_{AB}$, $\vert \phi_3\rangle^S_{AB}$, and $\vert \phi_4\rangle^S_{AB}$) and the even-parity mode (i.e., $\vert \phi_1\rangle^P_{AB}$, $\vert \phi_2\rangle^P_{AB}$, $\vert \phi_1\rangle^S_{AB}$, and $\vert \phi_2\rangle^S_{AB}$) of
the hyperentangled Bell states can be distinguished in both the
polarization and the spatial-mode DOFs, which is the result of the
polarization-spatial parity-check QND.

\section{ Two-step hyper-EPP for mixed hyperentangled Bell states with the quantum-state-joining method} \label{sec3}

\begin{figure}[!h]
\centering\includegraphics[width=8 cm,angle=0]{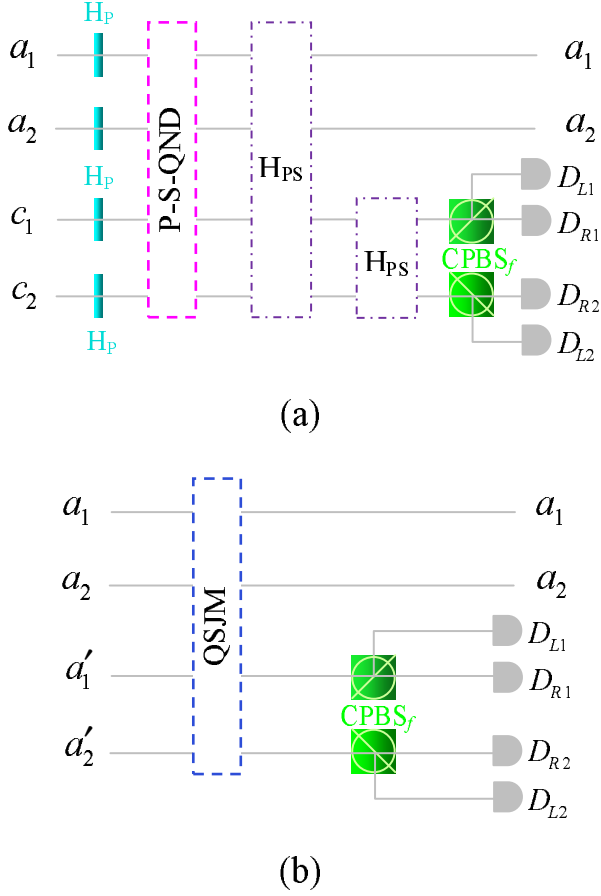} \caption{
(Color online) (a) Schematic diagram of the first step of our
hyper-EPP for mixed hyperentangled Bell states  with  polarization
bit-flip errors and spatial-mode phase-flip errors,  resorting to
the P-S-QND.  Bob performs the same operations as Alice by replacing
photons $A$ and $C$ with photons $B$ and $D$, respectively. (b)
Schematic diagram of the second step of our hyper-EPP for mixed
hyperentangled Bell states with polarization bit-flip errors and
spatial-mode phase-flip errors, resorting to the QSJM. Bob performs
the same operations as Alice by replacing photons $A$ and $A'$ with
photons $B$ and $B'$, respectively. $D_j$ ($j=L1$, $R1$, $R2$, or
$L2$) represents a single-photon detector.} \label{figure4}
\end{figure}

In this section, we introduce our two-step hyperentanglement
purification for mixed hyperentangled Bell states with polarization
bit-flip errors and spatial-mode phase-flip errors, resorting to the
P-S-QND and QSJM introduced in Sec. \ref{sec2}.  The setup of our
two-step hyper-EPP with QSJM is shown in Fig. \ref{figure4}. It
includes two steps in each round of the hyper-EPP process, and they
are discussed in detail as follows.

\subsection{ The first step of our two-step hyper-EPP with P-S-QNDs} \label{sec31}

In this step, two identical two-photon systems in a mixed
hyperentangled Bell state are required. They are described as
follows:
\begin{eqnarray}                           
\rho_{AB}&=&\left[F_1|\phi_1\rangle^P_{AB}\langle\phi_1|+(1-F_1)|\phi_3\rangle^P_{AB}\langle\phi_3|\right]\nonumber\\
&&\otimes\left[F_2|\phi_1\rangle^S_{AB}\langle\phi_1|+(1-F_2)|\phi_2\rangle^S_{AB}\langle\phi_2|\right],\nonumber\\
\rho_{CD}&=&\left[F_1|\phi_1\rangle^P_{CD}\langle\phi_1|+(1-F_1)|\phi_3\rangle^P_{CD}\langle\phi_3|\right]\nonumber\\
&&\otimes\left[F_2|\phi_1\rangle^S_{CD}\langle\phi_1|+(1-F_2)|\phi_2\rangle^S_{CD}\langle\phi_2|\right].\;\;\;\;
\end{eqnarray}
Here the subscripts $AB$ and $CD$ represent two photon pairs shared
by the two remote parties, called Alice and Bob, respectively.
Photons $A$ and $C$ belong to Alice, and photons $B$ and $D$ belong
to Bob.  $F_1$ and $F_2$ represent the probabilities of states
$|\phi_1\rangle^P_{AB}$ ($|\phi_1\rangle^P_{CD}$) and
$|\phi_1\rangle^S_{AB}$ ($|\phi_1\rangle^S_{CD}$) in the mixed
states, respectively.

The initial state of the four-photon system $ABCD$ is
$\rho_0=\rho_{AB}\otimes\rho_{CD}$. It can be viewed as the mixture
of 16 maximally hyperentangled pure states. That is, it is the
mixture of the four pure states in the polarization DOF:
$|\phi_1\rangle^P_{AB}\otimes|\phi_1\rangle^P_{CD}$ with the
probability $F_1^2$,
$|\phi_1\rangle^P_{AB}\otimes|\phi_3\rangle^P_{CD}$ with the
probability $F_1(1-F_1)$,
$|\phi_3\rangle^P_{AB}\otimes|\phi_1\rangle^P_{CD}$ with the
probability $F_1(1-F_1)$, and
$|\phi_3\rangle^P_{AB}\otimes|\phi_3\rangle^P_{CD}$ with the
probability $(1-F_1)^2$. It is also the mixture of the four pure
states in the spatial-mode DOF:
$|\phi_1\rangle^S_{AB}\otimes|\phi_1\rangle^S_{CD}$ with the
probability $F_2^2$,
$|\phi_1\rangle^S_{AB}\otimes|\phi_2\rangle^S_{CD}$ with the
probability $F_2(1-F_2)$,
$|\phi_2\rangle^S_{AB}\otimes|\phi_1\rangle^S_{CD}$ with the
probability $F_2(1-F_2)$, and
$|\phi_2\rangle^S_{AB}\otimes|\phi_2\rangle^S_{CD}$ with the
probability $(1-F_2)^2$.

The first step of our hyper-EPP for mixed hyperentangled Bell states
with polarization bit-flip errors and spatial-mode phase-flip errors
is shown in Fig.\ref{figure4}(a), which requires Alice and Bob
perform the Hadamard operations $H_P$ and P-S-QNDs on both the
polarization and the spatial-mode DOFs of photon pairs $AC$ and $BD$
in sequence. After performing $H_{PS}$ on the polarization and the
spatial-mode DOFs of photon pairs $AC$ and $BD$, Alice and Bob
measure the states of the excess electron spins in QDs to read out
the outcomes of the parity modes of photon pairs $AC$ and $BD$ in
both the polarization and the spatial-mode DOFs.

(1) If the two photon pairs, $AC$ and $BD$, are in the same
polarization parity mode and the same spatial-mode parity mode (both
in either the even-parity mode or the odd-parity mode), the
polarization pure states
$|\phi_1\rangle^P_{AB}\otimes|\phi_1\rangle^P_{CD}$ and
$|\phi_3\rangle^P_{AB}\otimes|\phi_3\rangle^P_{CD}$ are
distinguished from the other two polarization pure states,
$|\phi_1\rangle^P_{AB}\otimes|\phi_3\rangle^P_{CD}$ and
$|\phi_3\rangle^P_{AB}\otimes|\phi_1\rangle^P_{CD}$, and the
spatial-mode pure states
$|\phi_1\rangle^S_{AB}\otimes|\phi_1\rangle^S_{CD}$ and
$|\phi_2\rangle^S_{AB}\otimes|\phi_2\rangle^S_{CD}$ are
distinguished from the other two spatial-mode pure states,
$|\phi_1\rangle^S_{AB}\otimes|\phi_2\rangle^S_{CD}$ and
$|\phi_2\rangle^S_{AB}\otimes|\phi_1\rangle^S_{CD}$. After the
measurement on the excess electron spins in P-S-QNDs, the
polarization DOF of the four-photon system is projected into a mixed
state composed of states $|\Phi_1\rangle_P$ (or $|\Phi_2\rangle_P$)
and $|\Phi_3\rangle_P$ (or $|\Phi_4\rangle_P$), and the spatial-mode
DOF of the four-photon system is projected into a mixed state
composed of states $|\Phi_1\rangle_S$ (or $|\Phi_2\rangle_S$) and
$|\Phi_3\rangle_S$ (or $|\Phi_4\rangle_S$). Here
\begin{eqnarray}                           
|\Phi_1\rangle_P&=&\frac{1}{\sqrt{2}}(|RRRR\rangle+|LLLL\rangle)_{ABCD},\nonumber\\\
|\Phi_2\rangle_P&=&\frac{1}{\sqrt{2}}(|RRLL\rangle+|LLRR\rangle)_{ABCD},\nonumber\\\
|\Phi_3\rangle_P&=&\frac{1}{\sqrt{2}}(|RLRL\rangle+|LRLR\rangle)_{ABCD},\nonumber\\\
|\Phi_4\rangle_P&=&\frac{1}{\sqrt{2}}(|RLLR\rangle+|LRRL\rangle)_{ABCD},\nonumber\\\
|\Phi_1\rangle_S&=&\frac{1}{\sqrt{2}}(|a_1b_1c_1d_1\rangle+|a_2b_2c_2d_2\rangle),\nonumber\\\
|\Phi_2\rangle_S&=&\frac{1}{\sqrt{2}}(|a_1b_1c_2d_2\rangle+|a_2b_2c_1d_1\rangle),\nonumber\\\
|\Phi_3\rangle_S&=&\frac{1}{\sqrt{2}}(|a_1b_2c_1d_2\rangle+|a_2b_1c_2d_1\rangle),\nonumber\\\
|\Phi_4\rangle_S&=&\frac{1}{\sqrt{2}}(|a_1b_2c_2d_1\rangle+|a_2b_1c_1d_2\rangle).
\end{eqnarray}
State $|\Phi_2\rangle_P$ can be transformed into $|\Phi_1\rangle_P$
by performing the polarization bit-flip operations $\sigma_x^P$ on
both photons $C$ and $D$. In a similar way, states
$|\Phi_4\rangle_P$, $|\Phi_2\rangle_S$, and $|\Phi_4\rangle_S$ can
be transformed into states $|\Phi_3\rangle_P$, $|\Phi_1\rangle_S$,
and $|\Phi_3\rangle_S$, respectively. Subsequently, Alice and Bob
perform the Hadamard operations on both the polarization and the
spatial-mode DOFs of photons $C$ and $D$, and states
$|\Phi_1\rangle_P$, $|\Phi_3\rangle_P$, $|\Phi_1\rangle_S$, and
$|\Phi_3\rangle_S$ are transformed into states $|\Phi'_1\rangle_P$,
$|\Phi'_3\rangle_P$, $|\Phi'_1\rangle_S$, and $|\Phi'_3\rangle_S$,
respectively. Here
\begin{eqnarray}                           
|\Phi'_1\rangle_P\!\!&=&\!\!\frac{1}{2\sqrt{2}}[(|RR\rangle+|LL\rangle)_{AB}(|RR\rangle+|LL\rangle)_{CD}\nonumber\\
&&+(|RR\rangle-|LL\rangle)_{AB}(|RL\rangle+|LR\rangle)_{CD}],\nonumber\\\
|\Phi'_3\rangle_P\!\!&=&\!\!\frac{1}{2\sqrt{2}}[(|RL\rangle+|LR\rangle)_{AB}(|RR\rangle-|LL\rangle)_{CD}\nonumber\\
&&+(-|RL\rangle+|LR\rangle)_{AB}(|RL\rangle-|LR\rangle)_{CD}],\nonumber\\\
|\Phi'_1\rangle_S\!\!&=&\!\!\frac{1}{2\sqrt{2}}[(|a_1b_1\rangle+|a_2b_2\rangle)(|c_1d_1\rangle+|c_2d_2\rangle)\nonumber\\
&&+(|a_1b_1\rangle-|a_2b_2\rangle)(|c_1d_2\rangle+|c_2d_1\rangle)],\nonumber\\\
|\Phi'_3\rangle_S\!\!&=&\!\!\frac{1}{2\sqrt{2}}[(|a_1b_2\rangle+|a_2b_1\rangle)(|c_1d_1\rangle-|c_2d_2\rangle)\nonumber\\
&&+(-|a_1b_2\rangle+|a_2b_1\rangle)(|c_1d_2\rangle-|c_2d_1\rangle)].\;\;\;\;\;\;\;\;
\end{eqnarray}
Finally, photons $C$ and $D$ are detected with single-photon
detectors,  shown in Fig.\ref{figure4}(a). If the two clicked
single-photon detectors of photons $C$ and $D$ are in the
even-parity spatial mode (even-parity polarization mode), the
spatial-mode (polarization) DOF of the two-photon system $AB$ is
projected into state $|\phi_1\rangle^S_{AB}$ or
$|\phi_3\rangle^S_{AB}$ ($|\phi_1\rangle^P_{AB}$ or
$|\phi_3\rangle^P_{AB}$). If the outcome of the two clicked
detectors is in the odd-parity spatial mode (odd-parity polarization
mode), a phase-flip operation $\sigma^S_z=|b_1\rangle\langle b_1| -
|b_2\rangle\langle b_2|$ ($\sigma^P_z$) on the spatial-mode
(polarization) DOF of photon $B$ is required to obtain state
$|\phi_1\rangle^S_{AB}$ or $|\phi_3\rangle^S_{AB}$
($|\phi_1\rangle^P_{AB}$ or $|\phi_3\rangle^P_{AB}$).

(2) If photon pairs $AC$ and $BD$ are in different polarization
parity modes and different spatial-mode parity modes (one pair is in
the even-parity mode and the other is in the odd-parity mode), the
polarization pure states
$|\phi_1\rangle^P_{AB}\otimes|\phi_3\rangle^P_{CD}$ and
$|\phi_3\rangle^P_{AB}\otimes|\phi_1\rangle^P_{CD}$ are
distinguished from the other two polarization pure states,
$|\phi_1\rangle^P_{AB}\otimes|\phi_1\rangle^P_{CD}$ and
$|\phi_3\rangle^P_{AB}\otimes|\phi_3\rangle^P_{CD}$, and the
spatial-mode pure states
$|\phi_1\rangle^S_{AB}\otimes|\phi_2\rangle^S_{CD}$ and
$|\phi_2\rangle^S_{AB}\otimes|\phi_1\rangle^S_{CD}$ are
distinguished from the other two spatial-mode pure states,
$|\phi_1\rangle^S_{AB}\otimes|\phi_1\rangle^S_{CD}$ and
$|\phi_2\rangle^S_{AB}\otimes|\phi_2\rangle^S_{CD}$, respectively.
After the measurement on the excess electron spins in P-S-QNDs, the
polarization DOF of the four-photon system is projected into a mixed
state composed of states $|\Phi_5\rangle_P$ (or $|\Phi_6\rangle_P$)
and $|\Phi_7\rangle_P$ (or $|\Phi_8\rangle_P$), and the spatial-mode
DOF of the four-photon system is projected into a mixed state
composed of states $|\Phi_5\rangle_S$ (or $|\Phi_6\rangle_S$) and
$|\Phi_7\rangle_S$ (or $|\Phi_8\rangle_S$). Here
\begin{eqnarray}                           
|\Phi_5\rangle_P&=&\frac{1}{\sqrt{2}}(|RRRL\rangle+|LLLR\rangle)_{ABCD},\nonumber\\\
|\Phi_6\rangle_P&=&\frac{1}{\sqrt{2}}(|RRLR\rangle+|LLRL\rangle)_{ABCD},\nonumber\\\
|\Phi_7\rangle_P&=&\frac{1}{\sqrt{2}}(|RLRR\rangle+|LRLL\rangle)_{ABCD},\nonumber\\\
|\Phi_8\rangle_P&=&\frac{1}{\sqrt{2}}(|RLLL\rangle+|LRRR\rangle)_{ABCD},\nonumber\\\
|\Phi_5\rangle_S&=&\frac{1}{\sqrt{2}}(|a_1b_1c_1d_2\rangle+|a_2b_2c_2d_1\rangle),\nonumber\\\
|\Phi_6\rangle_S&=&\frac{1}{\sqrt{2}}(|a_1b_1c_2d_1\rangle+|a_2b_2c_1d_2\rangle),\nonumber\\\
|\Phi_7\rangle_S&=&\frac{1}{\sqrt{2}}(|a_1b_2c_1d_1\rangle+|a_2b_1c_2d_2\rangle),\nonumber\\\
|\Phi_8\rangle_S&=&\frac{1}{\sqrt{2}}(|a_1b_2c_2d_2\rangle+|a_2b_1c_1d_1\rangle).\;\;\;\;\;\;\;\;
\end{eqnarray}
As Alice and Bob cannot identify which photon pair, $AB$ or $CD$,
has the polarization bit-flip error (the spatial-mode phase-flip
error), they have to discard the two photon pairs in this case.

(3) If photon pairs $AC$ and $BD$ are in the same polarization
parity mode and different spatial-mode parity modes, the
polarization pure states
$|\phi_1\rangle^P_{AB}\otimes|\phi_1\rangle^P_{CD}$ and
$|\phi_3\rangle^P_{AB}\otimes|\phi_3\rangle^P_{CD}$ are
distinguished from the other two polarization pure states,
$|\phi_1\rangle^P_{AB}\otimes|\phi_3\rangle^P_{CD}$ and
$|\phi_3\rangle^P_{AB}\otimes|\phi_1\rangle^P_{CD}$, and the
spatial-mode pure states
$|\phi_1\rangle^S_{AB}\otimes|\phi_2\rangle^S_{CD}$ and
$|\phi_2\rangle^S_{AB}\otimes|\phi_1\rangle^S_{CD}$ are
distinguished from the other two spatial-mode pure states,
$|\phi_1\rangle^S_{AB}\otimes|\phi_1\rangle^S_{CD}$ and
$|\phi_2\rangle^S_{AB}\otimes|\phi_2\rangle^S_{CD}$. After the
measurement on the excess electron spins in P-S-QNDs, the
polarization DOF of the four-photon system is projected into a mixed
state composed of states $|\Phi_1\rangle_P$ (or $|\Phi_2\rangle_P$)
and $|\Phi_3\rangle_P$ (or $|\Phi_4\rangle_P$), and the spatial-mode
DOF of the four-photon system is projected into a mixed state
composed of states $|\Phi_5\rangle_S$ (or $|\Phi_6\rangle_S$) and
$|\Phi_7\rangle_S$ (or $|\Phi_8\rangle_S$). In this case, the second
step is required in our hyper-EPP with QSJM.

(4) If photon pairs $AC$ and $BD$ are in different polarization
parity modes and the same spatial-mode parity mode, the polarization
pure states $|\phi_1\rangle^P_{AB}\otimes|\phi_3\rangle^P_{CD}$ and
$|\phi_3\rangle^P_{AB}\otimes|\phi_1\rangle^P_{CD}$ are
distinguished from the other two polarization pure states,
$|\phi_1\rangle^P_{AB}\otimes|\phi_1\rangle^P_{CD}$ and
$|\phi_3\rangle^P_{AB}\otimes|\phi_3\rangle^P_{CD}$, and the
spatial-mode pure states
$|\phi_1\rangle^S_{AB}\otimes|\phi_1\rangle^S_{CD}$ and
$|\phi_2\rangle^S_{AB}\otimes|\phi_2\rangle^S_{CD}$ are
distinguished from the other two spatial-mode pure states,
$|\phi_1\rangle^S_{AB}\otimes|\phi_2\rangle^S_{CD}$ and
$|\phi_2\rangle^S_{AB}\otimes|\phi_1\rangle^S_{CD}$. After the
measurement on the excess electron spins in P-S-QNDs, the
polarization DOF of the four-photon system is projected into a mixed
state composed of states $|\Phi_5\rangle_P$ ( or $|\Phi_6\rangle_P$)
and $|\Phi_7\rangle_P$ (or $|\Phi_8\rangle_P$), and the spatial-mode
DOF of the four-photon system is projected into a mixed state
composed of states $|\Phi_1\rangle_S$ (or $|\Phi_2\rangle_S$) and
$|\Phi_3\rangle_S$ (or $|\Phi_4\rangle_S$). In this case, the second
step is required in our hyper-EPP with QSJM as well.

\subsection{ The second step of our two-step hyper-EPP with QSJM} \label{sec32}

In this step, we show that the efficiency of the hyper-EPP is
improved with  QSJM by preserving cases (3) and (4) in the first
step, as shown in Fig. \ref{figure4}(b).

Suppose that there are four identical photon pairs, $AB$, $CD$,
$A'B'$, and $C'D'$, shared by Alice and Bob. Photons $A$, $C$, $A'$,
$C'$ belong to Alice, and photons $B$, $D$, $B'$, $D'$ belong to
Bob. Alice and Bob perform the same operations on the four-photon
systems $ABCD$ and $A'B'C'D'$ as they did in the first step. If the
four-photon systems $ABCD$ and $A'B'C'D'$ are projected into the
states in cases (4) and (3) in the first step, respectively, Alice
and Bob can use the QSJM (introduced in Sec. \ref{sec22}) to
transfer the polarization state of the four-photon system $A'B'C'D'$
to the polarization state of the four-photon system $ABCD$
(discussed in Appendix \ref{appendixa}). The polarization state of
the four-photon system ABCD is in a mixed state composed of states
$|\Phi_1\rangle_P$ and $|\Phi_3\rangle_P$,  and the spatial-mode DOF
of the four-photon system ABCD is in a mixed state composed of
states $|\Phi_1\rangle_S$ and $|\Phi_3\rangle_S$, which is the
preserving condition of case (1) in the first step. After Hadamard
operations and detections are performed on both the polarization and
spatial-mode DOFs of photons $C$ and $D$ and the conditional
phase-flip operation $\sigma^S_z$ ($\sigma^P_z$) is performed on
photon $B$, the spatial-mode states $|\phi_1\rangle^S_{AB}$ and
$|\phi_3\rangle^S_{AB}$ (polarization states $|\phi_1\rangle^P_{AB}$
and $|\phi_3\rangle^P_{AB}$) are obtained.

If the four-photon systems $ABCD$ and $A'B'C'D'$ are projected into
the states in cases (3) and (4) in the first step, respectively,
Alice and Bob can use the QSJM to transfer the spatial-mode state of
the four-photon system $A'B'C'D'$ to the spatial-mode state of the
four-photon system $ABCD$. Then the preserving condition of case (1)
in the first step is achieved. After Hadamard operations and
detections are performed on both the polarization and spatial-mode
DOFs of photons $C$ and $D$ and the conditional phase-flip operation
$\sigma^S_z$ ($\sigma^P_z$) is performed on photon $B$, the
spatial-mode states $|\phi_1\rangle^S_{AB}$ and
$|\phi_3\rangle^S_{AB}$ (polarization states $|\phi_1\rangle^P_{AB}$
and $|\phi_3\rangle^P_{AB}$) are obtained.

\begin{figure}[htb]                    
\centering
\includegraphics[width=8 cm]{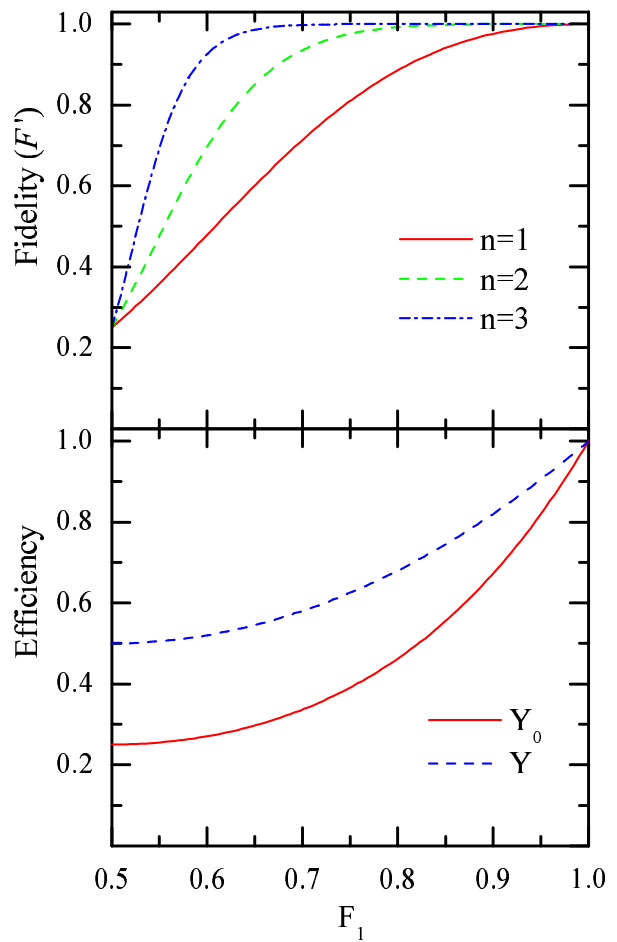}
\caption{(Color online) The fidelity and efficiency of our hyper-EPP
for mixed hyperentangled Bell states. $F'=F'_1\times F'_2$. $n$ is
the iteration number of the hyper-EPP process. $Y_0$ and $Y$
represent the efficiencies of the first round of the hyper-EPP
processes without and with  QSJM, respectively. The parameters of the
mixed hyperentangled Bell state are chosen to be $F_1=F_2$.}
\label{figure5}
\end{figure}

After the first round of our hyper-EPP process with these two steps,
the state of the photon pair $AB$ is transformed into
\begin{eqnarray}                           
\rho'_{AB}&=&\left[F'_1|\phi_1\rangle^P_{AB}\langle\phi_1|+(1-F'_1)|\phi_3\rangle^P_{AB}\langle\phi_3|\right] \nonumber\\
&&\otimes\left[F'_2|\phi_1\rangle^S_{AB}\langle\phi_1|+(1-F'_2)|\phi_3\rangle^S_{AB}\langle\phi_3|\right].\;\;\;\;\;\;\;\;\;\label{eq15}
\end{eqnarray}
Here $F'_1=\frac{F_1^2}{[F_1^2+(1-F_1)^2]}$,
$F'_2=\frac{F_2^2}{[F_2^2+(1-F_2)^2]}$, and $F_i>1/2$ ($i=1,2$). The
fidelity of state
$|\phi_1\rangle^P_{AB}\otimes|\phi_1\rangle^S_{AB}$ in Eq.
(\ref{eq15}) is $F'=F'_1\times F'_2$. State $|\phi_3\rangle^S_{AB}$
can be transformed into $|\phi_2\rangle^S_{AB}$ with the Hadamard
operations on the spatial-mode DOF of photons $A$ and $B$. With the
iteration of our hyper-EPP process, the fidelity of the two-photon
state can be improved (shown in Fig. \ref{figure5} for the cases
with $F_1=F_2$).

The efficiency of an EPP is defined as the probability of obtaining
a high-fidelity entangled photon system from a pair of photon
systems transmitted over a noisy channel without photon loss. Our
hyper-EPP is constructed to purify the mixed hyperentangled Bell
states with spatial-mode phase-flip errors and polarization bit-flip
errors. In our previous work \cite{HEPP}, we introduced a hyper-EPP
for the mixed hyperentangled Bell states with bit-flip errors in
both the spatial-mode and polarization DOFs, which only preserves
case (1) in the first step with the probability of
\begin{eqnarray}                           
\left[F_1^2+(1-F_1)^2\right]\left[F_2^2+(1-F_2)^2\right]
\end{eqnarray}
(in the first round of the hyper-EPP process), and the efficiency is
\begin{eqnarray}                           
Y_0=\left[F_1^2+(1-F_1)^2\right]\left[F_2^2+(1-F_2)^2\right].
\end{eqnarray}
In the present hyper-EPP, cases (3) and (4) in the first step can
also be preserved with our QSJM, and the efficiency of the first
round of the hyper-EPP process is
\begin{eqnarray}                           
Y=F_2^2\!+\!(1\!-\!F_2)^2
\end{eqnarray}
when $F_1>F_2$. The efficiency of the hyper-EPP is improved by
resorting to the QSJM (shown in Fig. \ref{figure5} for the cases
with $F_1=F_2$).

\section{Hyper-EPP for hyperentangled GHZ states with  QSJM} \label{sec4}

We can generalize our hyper-EPP for mixed hyperentangled GHZ states
of three-photon systems. There are eight polarization GHZ states and
eight spatial-mode GHZ states, and they are described as follows:
\begin{eqnarray}                           
|\psi^\pm_0\rangle_P&=&\frac{1}{\sqrt{2}}(|RRR\rangle\pm|LLL\rangle)_{ABC},\nonumber\\\
|\psi^\pm_1\rangle_P&=&\frac{1}{\sqrt{2}}(|RRL\rangle\pm|LLR\rangle)_{ABC},\nonumber\\\
|\psi^\pm_2\rangle_P&=&\frac{1}{\sqrt{2}}(|RLR\rangle\pm|LRL\rangle)_{ABC},\nonumber\\\
|\psi^\pm_3\rangle_P&=&\frac{1}{\sqrt{2}}(|LRR\rangle\pm|RLL\rangle)_{ABC},\nonumber\\\
|\psi^\pm_0\rangle_S&=&\frac{1}{\sqrt{2}}(|a_1b_1c_1\rangle\pm|a_2b_2c_2\rangle),\nonumber\\\
|\psi^\pm_1\rangle_S&=&\frac{1}{\sqrt{2}}(|a_1b_1c_2\rangle\pm|a_2b_2c_1\rangle),\nonumber\\\
|\psi^\pm_2\rangle_S&=&\frac{1}{\sqrt{2}}(|a_1b_2c_1\rangle\pm|a_2b_1c_2\rangle),\nonumber\\\
|\psi^\pm_3\rangle_S&=&\frac{1}{\sqrt{2}}(|a_2b_1c_1\rangle\pm|a_1b_2c_2\rangle).
\end{eqnarray}
Here the subscripts $A$, $B$, and $C$ represent the photons obtained
by the remote users Alice, Bob, and Charlie, respectively. Suppose
that the original state of the three-photon system $ABC$ is
$|\psi^+_0\rangle_P\otimes|\psi^+_0\rangle_S$. If there is a
polarization bit-flip error on the original state of the
three-photon system after its transmission over a noisy channel, the
polarization state of the three-photon system will become
$|\psi^+_1\rangle_P$, $|\psi^+_2\rangle_P$, or $|\psi^+_3\rangle_P$.
However, a spatial-mode phase-flip error is more likely to take
place on the original state of the three-photon system
\cite{HEPP,LI}, and the spatial-mode state of the three-photon
system will become $|\psi^-_0\rangle_S$.

After the photon system is transmitted over the noisy channel with
polarization bit-flip errors and spatial-mode phase-flip errors, its
state  becomes
\begin{eqnarray}                           
\rho&=&\big(F_0|\psi^+_0\rangle_P\langle\psi^+_0|+F_1|\psi^+_1\rangle_P\langle\psi^+_1|\nonumber\\
&&+F_2|\psi^+_2\rangle_P\langle\psi^+_2|+F_3|\psi^+_3\rangle_P\langle\psi^+_3|\big)\nonumber\\
&&\otimes\big(P_0|\psi^+_0\rangle_S\langle\psi^+_0|+P_1|\psi^-_0\rangle_S\langle\psi^-_0|\big).
\;\;\;\;\;\;\;\;\;\label{eq17}
\end{eqnarray}
Here $F_0$ and $P_0$ are the probabilities of states
$|\psi^+_0\rangle_P$ and $|\psi^+_0\rangle_S$ in the mixed states,
respectively, and they satisfy the relations $F_0+F_1+F_2+F_3=1$ and
$P_0+P_1=1$. In order to obtain high-fidelity entangled three-photon
systems, the three remote users, Alice, Bob, and Charlie, have to
perform hyper-EPP on the three-photon systems, which requires them
to divide their photon systems into many groups with a pair of
three-photon systems in each group. The photons in each group are
labeled $ABCA'B'C'$, where $ABC$ and $A'B'C'$ represent two
identical three-photon systems in the same mixed hyperentangled GHZ
state. As the phase-flip error of the three-photon  GHZ state cannot
be transformed into the bit-flip error with Hadamard operations, we
discuss the principles of the EPPs for the polarization states and
the spatial-mode states of the three-photon systems in Secs.
\ref{sec41} and \ref{sec42}, respectively. The hyper-EPP for mixed
hyperentangled GHZ states with our QSJM is introduced in Sec.
\ref{sec43}.

\subsection{EPP for polarization GHZ states}
\label{sec41}

The polarization state of system $ABCA'B'C'$ can be viewed as the
mixture of 16 pure states
$|\psi^+_i\rangle_P\otimes|\psi^+_j\rangle_P$ with the probability
of $F_iF_j$ ($i,j=0,1,2,3$). Alice, Bob, and Charlie perform
polarization parity-check QNDs on their photon pairs $AA'$, $BB'$,
and $CC'$, respectively, and they pick the groups in which the three
photon pairs are all in the even-parity polarization mode or all in
the odd-parity polarization mode. If the polarization states of the
three photon pairs are all in the even-parity mode, the polarization
DOF of the six-photon system will project into a mixed state
composed of the four pure states $|\Psi_0\rangle_P$,
$|\Psi_1\rangle_P$, $|\Psi_2\rangle_P$, and $|\Psi_3\rangle_P$. Here
\begin{eqnarray}                           
|\Psi_0\rangle_P\!\!\!&=&\!\!\!\frac{1}{\sqrt{2}}(|RRR\rangle|RRR\rangle+|LLL\rangle|LLL\rangle)_{ABCA'B'C'},\nonumber\\\
|\Psi_1\rangle_P\!\!\!&=&\!\!\!\frac{1}{\sqrt{2}}(|RRL\rangle|RRL\rangle+|LLR\rangle|LLR\rangle)_{ABCA'B'C'},\nonumber\\\
|\Psi_2\rangle_P\!\!\!&=&\!\!\!\frac{1}{\sqrt{2}}(|RLR\rangle|RLR\rangle+|LRL\rangle|LRL\rangle)_{ABCA'B'C'},\nonumber\\\
|\Psi_3\rangle_P\!\!\!&=&\!\!\!\frac{1}{\sqrt{2}}(|LRR\rangle|LRR\rangle+|RLL\rangle|RLL\rangle)_{ABCA'B'C'}.\nonumber\\\
\end{eqnarray}

If the polarization states of the three photon pairs are all in the
odd-parity mode, the polarization DOF of the six-photon system will
project into a mixed state composed of the four pure states
$|\Psi'_0\rangle_P$, $|\Psi'_1\rangle_P$, $|\Psi'_2\rangle_P$, and
$|\Psi'_3\rangle_P$. Here
\begin{eqnarray}                           
|\Psi'_0\rangle_P\!\!\!&=&\!\!\!\frac{1}{\sqrt{2}}(|RRR\rangle|LLL\rangle+|LLL\rangle|RRR\rangle)_{ABCA'B'C'},\nonumber\\\
|\Psi'_1\rangle_P\!\!\!&=&\!\!\!\frac{1}{\sqrt{2}}(|RRL\rangle|LLR\rangle+|LLR\rangle|RRL\rangle)_{ABCA'B'C'},\nonumber\\\
|\Psi'_2\rangle_P\!\!\!&=&\!\!\!\frac{1}{\sqrt{2}}(|RLR\rangle|LRL\rangle+|LRL\rangle|RLR\rangle)_{ABCA'B'C'},\nonumber\\\
|\Psi'_3\rangle_P\!\!\!&=&\!\!\!\frac{1}{\sqrt{2}}(|LRR\rangle|RLL\rangle+|RLL\rangle|LRR\rangle)_{ABCA'B'C'}.\nonumber\\\
\end{eqnarray}

With the polarization bit-flip operations $\sigma_x^P$ on photons
$A'$, $B'$ and $C'$, the polarization state $|\Psi'_i\rangle_P$ can
be transformed into $|\Psi_i\rangle_P$ ($i=0,1,2,3$). After Hardmard
operations and detections are performed on the polarization DOF of
photons $A'$, $B'$ and $C'$ and the conditional phase-flip operation
$\sigma^P_z$ is performed on photon $A$ \cite{MEPP2}, states
$|\psi^+_0\rangle_P$, $|\psi^+_1\rangle_P$, $|\psi^+_2\rangle_P$,
and $|\psi^+_3\rangle_P$ are obtained with the probabilities
$F_0^2$, $F_1^2$, $F_2^2$, and $F_3^2$, respectively. If the results
of the three polarization parity-check QNDs have other outcomes,
which of the three-photon systems, $ABC$ or $A'B'C'$, has the
polarization bit-flip error is ambiguous, which will be discussed in
the hyper-EPP with our QSJM in Sec. \ref{sec43}.

\subsection{EPP for spatial-mode GHZ states}
\label{sec42}

The spatial-mode state of system $ABCA'B'C'$ can be viewed as the
mixture of four pure states
$|\psi^+_0\rangle_S\otimes|\psi^+_0\rangle_S$,
$|\psi^+_0\rangle_S\otimes|\psi^-_0\rangle_S$,
$|\psi^-_0\rangle_S\otimes|\psi^+_0\rangle_S$, and
$|\psi^-_0\rangle_S\otimes|\psi^-_0\rangle_S$, with the
probabilities of $P^2_0$, $P_0P_1$, $P_0P_1$, and $P^2_1$,
respectively. With spatial-mode Hadamard operations on the three
photons in each subsystem, states $|\psi^+_0\rangle_S$ and
$|\psi^-_0\rangle_S$ are transformed into states
$|\varphi^+\rangle_S$ and $|\varphi^-\rangle_S$, respectively. Here
\begin{eqnarray}                           
|\varphi^+\rangle_S&=&\frac{1}{2}(|a_1b_1c_1\rangle+|a_1b_2c_2\rangle+|a_2b_2c_1\rangle+|a_2b_1c_2\rangle),\nonumber\\\
|\varphi^-\rangle_S&=&\frac{1}{2}(|a_1b_1c_2\rangle+|a_1b_2c_1\rangle+|a_2b_2c_2\rangle+|a_2b_1c_1\rangle).\nonumber\\\
\end{eqnarray}
Alice, Bob, and Charlie perform spatial-mode parity-check QNDs on
their photon pairs $AA'$, $BB'$, and $CC'$, respectively, and they
pick the groups with an odd number of photon pairs in the
even-parity spatial mode.

If the three photon pairs are all in the even-parity spatial mode,
the spatial-mode DOF of the six-photon system is projected  into a
mixed state composed of the two pure states $|\Psi_0\rangle_S$ and
$|\Psi_1\rangle_S$. Here
\begin{eqnarray}                           
|\Psi_0\rangle_S&=&\frac{1}{2}(|a_1b_1c_1\rangle|a'_1b'_1c'_1\rangle+|a_1b_2c_2\rangle|a'_1b'_2c'_2\rangle\nonumber\\\
&&+|a_2b_2c_1\rangle|a'_2b'_2c'_1\rangle+|a_2b_1c_2\rangle|a'_2b'_1c'_2\rangle),\nonumber\\\
|\Psi_1\rangle_S&=&\frac{1}{2}(|a_1b_1c_2\rangle|a'_1b'_1c'_2\rangle+|a_1b_2c_1\rangle|a'_1b'_2c'_1\rangle\nonumber\\\
&&+|a_2b_2c_2\rangle|a'_2b'_2c'_2\rangle+|a_2b_1c_1\rangle|a'_2b'_1c'_1\rangle).\;\;\;\;\;\;\;\;
\end{eqnarray}

If photon pairs $AA'$ and $BB'$ are in the odd-parity spatial mode
and photon pair $CC'$ is in the even-parity spatial mode, the
spatial-mode DOF of the six-photon system is projected into a mixed
state composed of the two pure states $|\Psi'_0\rangle_S$ and
$|\Psi'_1\rangle_S$. Here
\begin{eqnarray}                           
|\Psi'_0\rangle_S&=&\frac{1}{2}(|a_1b_1c_1\rangle|a'_2b'_2c'_1\rangle+|a_2b_2c_1\rangle|a'_1b'_1c'_1\rangle\nonumber\\\
&&+|a_1b_2c_2\rangle|a'_2b'_1c'_2\rangle+|a_2b_1c_2\rangle|a'_1b'_2c'_2\rangle),\nonumber\\\
|\Psi'_1\rangle_S&=&\frac{1}{2}(|a_1b_1c_2\rangle|a'_2b'_2c'_2\rangle+|a_2b_2c_2\rangle|a'_1b'_1c'_2\rangle\nonumber\\\
&&+|a_1b_2c_1\rangle|a'_2b'_1c'_1\rangle+|a_2b_1c_1\rangle|a'_1b'_2c'_1\rangle).\;\;\;\;\;\;\;\;
\end{eqnarray}

If photon pairs $AA'$ and $CC'$ are in  the  odd-parity spatial mode
and photon pair $BB'$ is in  the  even-parity spatial mode, the
spatial-mode DOF of the six-photon system is projected into a mixed
state composed of the  two pure states $|\Psi'_2\rangle_S$ and
$|\Psi'_3\rangle_S$. Here
\begin{eqnarray}                           
|\Psi'_2\rangle_S&=&\frac{1}{2}(|a_1b_1c_1\rangle|a'_2b'_1c'_2\rangle+|a_2b_2c_1\rangle|a'_1b'_2c'_2\rangle\nonumber\\\
&&+|a_1b_2c_2\rangle|a'_2b'_2c'_1\rangle+|a_2b_1c_2\rangle|a'_1b'_1c'_1\rangle),\nonumber\\\
|\Psi'_3\rangle_S&=&\frac{1}{2}(|a_1b_1c_2\rangle|a'_2b'_1c'_1\rangle+|a_2b_2c_2\rangle|a'_1b'_2c'_1\rangle\nonumber\\\
&&+|a_1b_2c_1\rangle|a'_2b'_2c'_2\rangle+|a_2b_1c_1\rangle|a'_1b'_1c'_2\rangle).\;\;\;\;\;\;\;
\end{eqnarray}

If photon pairs $BB'$ and $CC'$ are in the odd-parity spatial mode
and photon pair $AA'$ is in the even-parity spatial mode, the
spatial-mode DOF of the six-photon system will project into a mixed
state composed of the two pure states $|\Psi'_4\rangle_S$ and
$|\Psi'_5\rangle_S$. Here
\begin{eqnarray}                           
|\Psi'_4\rangle_S&=&\frac{1}{2}(|a_1b_1c_1\rangle|a'_1b'_2c'_2\rangle+|a_2b_2c_1\rangle|a'_2b'_1c'_2\rangle\nonumber\\\
&&+|a_1b_2c_2\rangle|a'_1b'_1c'_1\rangle+|a_2b_1c_2\rangle|a'_2b'_2c'_1\rangle),\nonumber\\\
|\Psi'_5\rangle_S&=&\frac{1}{2}(|a_1b_1c_2\rangle|a'_1b'_2c'_1\rangle+|a_2b_2c_2\rangle|a'_2b'_1c'_1\rangle\nonumber\\\
&&+|a_1b_2c_1\rangle|a'_1b'_1c'_2\rangle+|a_2b_1c_1\rangle|a'_2b'_2c'_2\rangle).\;\;\;\;\;\;\;\;
\end{eqnarray}

With the spatial-mode bit-flip operations $\sigma_x^S$ on the two
photons of the three-photon system $A'B'C'$, the spatial-mode states
$|\Psi'_0\rangle_S$ (or $|\Psi'_2\rangle_S$, $|\Psi'_4\rangle_S$)
and $|\Psi'_1\rangle_S$ (or $|\Psi'_3\rangle_S$,
$|\Psi'_5\rangle_S$) can be transformed into $|\Psi_0\rangle_S$ and
$|\Psi_1\rangle_S$, respectively. After Hadamard operations and
detections are performed on the spatial-mode DOF of photons $A'B'C'$
and the conditional phase-flip operation $\sigma^S_z$ is performed
on photon $A$ (or $B$, $C$) \cite{MEPP2}, states
$|\varphi^+\rangle_S$ and $|\varphi^-\rangle_S$ are obtained with
the probabilities $P_0^2$ and $P_1^2$, respectively. If the results
of the three spatial-mode parity-check QNDs have other outcomes,
which of the three-photon systems, $ABC$ or $A'B'C'$, has the
spatial-mode phase-flip error is ambiguous, which will be discussed
in the hyper-EPP with our QSJM in Sec. \ref{sec43}. States
$|\varphi^+\rangle_S$ and $|\varphi^-\rangle_S$ can be transformed
into $|\psi^+_0\rangle_S$ and $|\psi^-_0\rangle_S$, respectively, by
performing spatial-mode Hadamard operations on the three-photon
systems.

\subsection{Hyper-EPP for hyperentangled GHZ states with QSJM}
\label{sec43}

\begin{figure}[htb]                    
\centering
\includegraphics[width=8 cm]{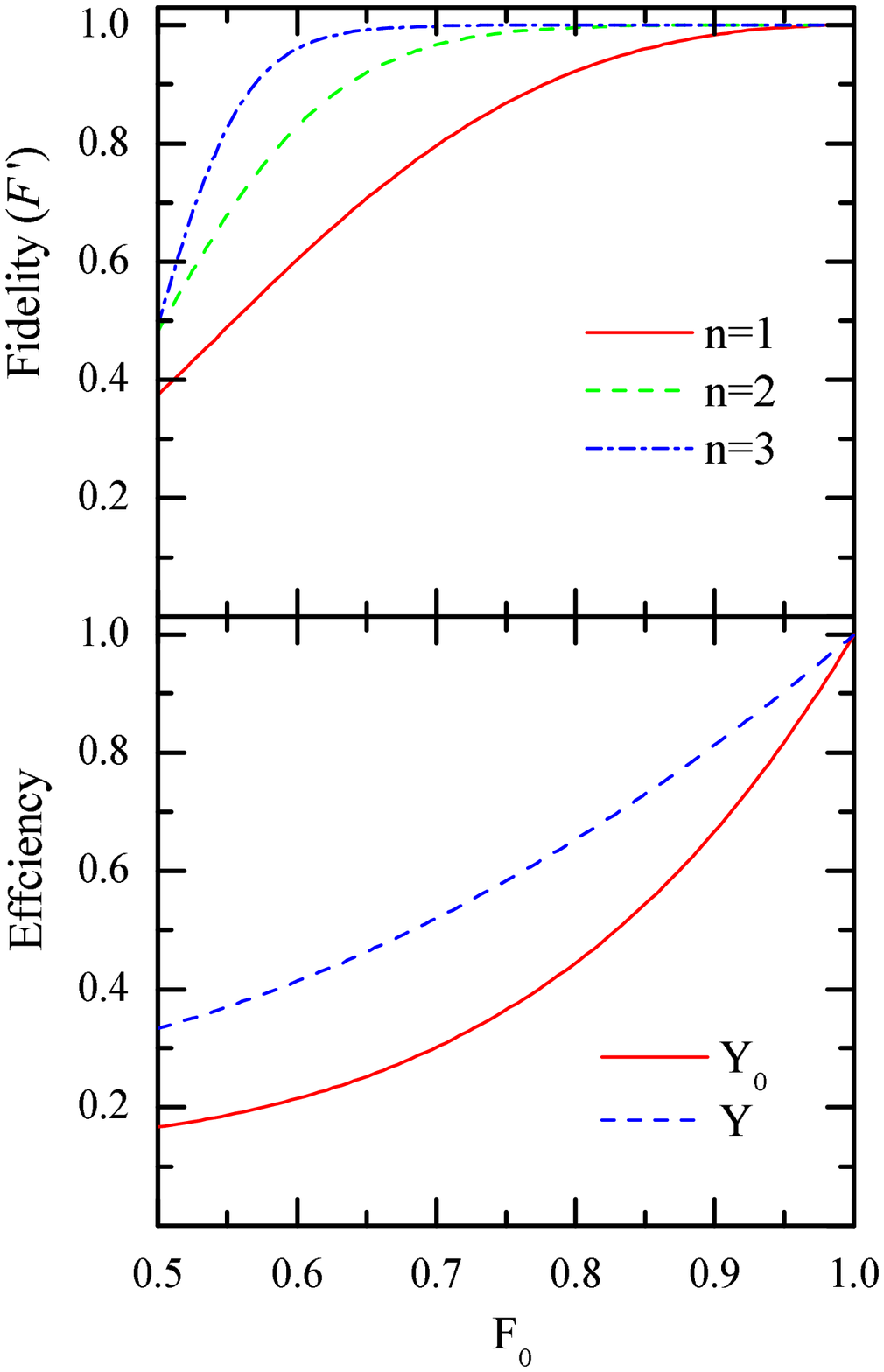}
\caption{ (Color online) The fidelity and the efficiency of our
hyper-EPP for mixed hyperentangled GHZ states. $F'=F'_0P'_0$. $n$ is
the iteration number of the hyper-EPP process. $Y_0$ and $Y$
represent the efficiencies of the first round of the hyper-EPP
processes without and with QSJM, respectively. The parameters of the
mixed hyperentangled GHZ state are chosen to be $F_1=F_2=F_3$ and
$P_0=F_0$.} \label{figure6}
\end{figure}

The hyper-EPP for mixed hyperentangled GHZ states with our QSJM can
also be implemented with the quantum circuit shown in Fig.
\ref{figure4}. Alice, Bob, and Charlie first perform the same
operations [shown in Fig. \ref{figure4}(a))] on their photon pairs
$AA'$, $BB'$, and $CC'$, respectively. If the results of the
P-S-QNDs in a group show that the three photon pairs are all in the
even- or odd-parity polarization mode and an odd number of photon
pairs are in the even-parity spatial mode, the group of the
three-photon systems is preserved.

If the three photon pairs are all in the even- or odd-parity
polarization mode and an even number of photon pairs are in the
even-parity spatial mode in a group, the polarization state of the
group can be transferred to the polarization state of another group
in which the three photon pairs are not all in the even- or
odd-parity polarization mode and an odd number of photon pairs are
in the even-parity spatial mode, resorting to the QSJM shown in Fig.
\ref{figure4}(b). Then the polarization DOF of the six-photon system
$ABCA'B'C'$ is in a mixed state composed of the four pure states
$|\Psi_0\rangle_P$, $|\Psi_1\rangle_P$, $|\Psi_2\rangle_P$, and
$|\Psi_3\rangle_P$, and the spatial-mode DOF of the six-photon
system $ABCA'B'C'$ is in a mixed state composed of the two pure
states $|\Psi_0\rangle_S$ and $|\Psi_1\rangle_S$, which is the
preserving condition in the first step.

If the three photon pairs are not all in the even- or odd-parity
polarization mode and an odd number of photon pairs are in the
even-parity spatial mode in a group, the spatial-mode state of the
group can be transferred  to the spatial-mode state of another group
in which the three photon pairs are all in the even- or odd-parity
polarization mode and an even number of photon pairs are in the
even-parity spatial mode, resorting to the QSJM. Then the preserving
condition in the first step is achieved. Finally, the group of the
three-photon systems has to be discarded if the results of the
P-S-QNDs show that the three photon pairs are not all in the even-
or odd-parity polarization mode and an even number of photon pairs
are in the even-parity spatial mode.

After the first round of this hyper-EPP process with the two steps,
the state of three-photon system $ABC$ is transformed into
\begin{eqnarray}                           
\rho&=&\big(F'_0|\psi^+_0\rangle_P\langle\psi^+_0|+F'_1|\psi^+_1\rangle_P\langle\psi^+_1|\nonumber\\
&&+F'_2|\psi^+_2\rangle_P\langle\psi^+_2|+F'_3|\psi^+_3\rangle_P\langle\psi^+_3|\big)\nonumber\\
&&\otimes\big(P'_0|\psi^+_0\rangle_S\langle\psi^+_0|+P'_1|\psi^-_0\rangle_S\langle\psi^-_0|\big).\;\;\;\;\;\;\;\;\;\label{eq25}
\end{eqnarray}
Here
\begin{eqnarray}                           
F'_0&=&\frac{F_0^2}{F_0^2+F_1^2+F_2^2+F_3^2},\nonumber\\
F'_1&=&\frac{F_1^2}{F_0^2+F_1^2+F_2^2+F_3^2},\nonumber\\
F'_2&=&\frac{F_2^2}{F_0^2+F_1^2+F_2^2+F_3^2},\nonumber\\
F'_3&=&\frac{F_3^2}{F_0^2+F_1^2+F_2^2+F_3^2},\nonumber\\
P'_0&=&\frac{P_0^2}{P_0^2+P_1^2},\nonumber\\
P'_1&=&\frac{P_1^2}{P_0^2+P_1^2}.\;\;\;\;\;\;\;
\end{eqnarray}
If $P_0>1/2$, the fidelity  $P'_0>P_0$. The fidelity $F'_0>F_0$ if
$F_0$ satisfies the relation
\begin{eqnarray}                           
F_0&>&\frac{1}{4}[3-2F_1-2F_2\nonumber\\
&&-\sqrt{1+4(F_1+F_2)-12(F_1^2+F_2^2)-8F_1F_2}].\;\;\;\;\;\;\;\;
\end{eqnarray}
The fidelity of $|\psi_0^+\rangle_P\otimes|\psi_0^+\rangle_S$ in Eq.
(\ref{eq25}) is $F'=F'_0P'_0$. With the iteration of our hyper-EPP
process, the fidelity of this three-photon state can be improved
(shown in Fig. \ref{figure6} for cases with $F_1=F_2=F_3$ and
$P_0=F_0$).

In the hyper-EPP without QSJM, the groups of three-photon systems
can only be preserved with the probability of
$(F_0^2+F_1^2+F_2^2+F_3^2)(P_0^2+P_1^2)$ in the first round of the
hyper-EPP process, and the efficiency is
\begin{eqnarray}                           
Y_0=(F_0^2+F_1^2+F_2^2+F_3^2)(P_0^2+P_1^2).
\end{eqnarray}
In this hyper-EPP with our QSJM, the probability of the preserved
three-photon systems in the first round of the hyper-EPP process can
be increased to
\begin{eqnarray}                           
\min\{(F_0^2+F_1^2+F_2^2+F_3^2),(P_0^2+P_1^2)\},
\end{eqnarray}
and the efficiency is
\begin{eqnarray}                           
Y&\!=\!&\min\{(P_0^2+P_1^2),(F_0^2+F_1^2+F_2^2+F_3^2)\}.\;\;\;\;\;\;\;\;
\end{eqnarray}
Figure \ref{figure6} shows the efficiencies of the hyper-EPPs (in
the first round) with and without QSJM (for the cases with
$F_1=F_2=F_3$ and $P_0=F_0$), and it is obvious that the efficiency
of the hyper-EPP with QSJM is greatly improved, compared with the
one without QSJM.

\section{Discussion and summary}
\label{sec5}

The transmission and reflection rule of the double-sided QD-cavity
system is the key element for the P-S-QND and the QSJM, and it may
not be perfect because of decoherence and dephasing.  The electron
spin decoherence may reduce the fidelity of the proposal, but this
effect can be suppressed by extending the electron coherence time to
microseconds using spin-echo techniques (using single-photon pluses to
play the role of the $\pi$ pulse), which is longer than the cavity
photon lifetime ($\sim10p$s) and the time interval of the input
photons (approximately nanoseconds)) \cite{QD5}. The exciton dephasing, including
optical dephasing and hole spin dephasing, can also reduce the
fidelity by a few percent. The hole spin coherence time is three
times order of magnitude longer than the cavity photon lifetime
\cite{trion5,trion6}, and the optical coherence time of the exciton
($\sim100p$s) is ten times longer than the cavity photon lifetime
\cite{trion1,trion2}, so the effect of the exciton dephasing may be
decreased. The heavy-light hole mixing may also reduce the fidelity
by a few percent, but this effect can be decreased by improving the
shape, size, and type of QDs \cite{QD5}. The fine-structure
splitting could be unaffected by the charged exciton due to the quenched
exchange interaction \cite{fine,fine1}. Also, the Hadamard operation,
which is used to transform electron-spin states $|\uparrow\rangle$
and $|\downarrow\rangle$ into $|+\rangle$ and $|-\rangle$,
respectively, can be implemented by nanosecond electron-spin-resonance
microwave pulses or picosecond optical pulses \cite{manu}.

\begin{figure}[htb]                    
\centering
\includegraphics[width=7 cm]{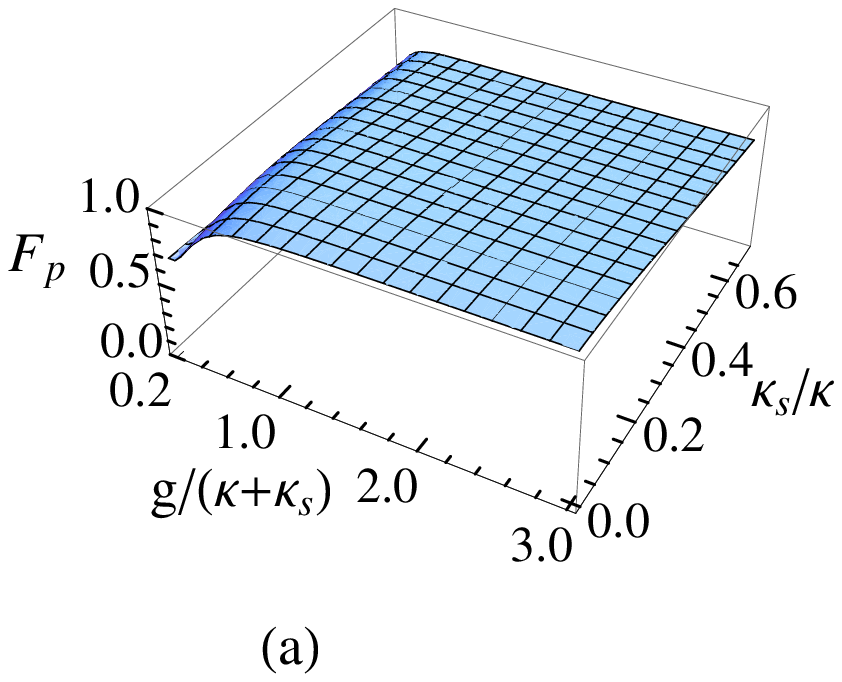}
\includegraphics[width=7 cm]{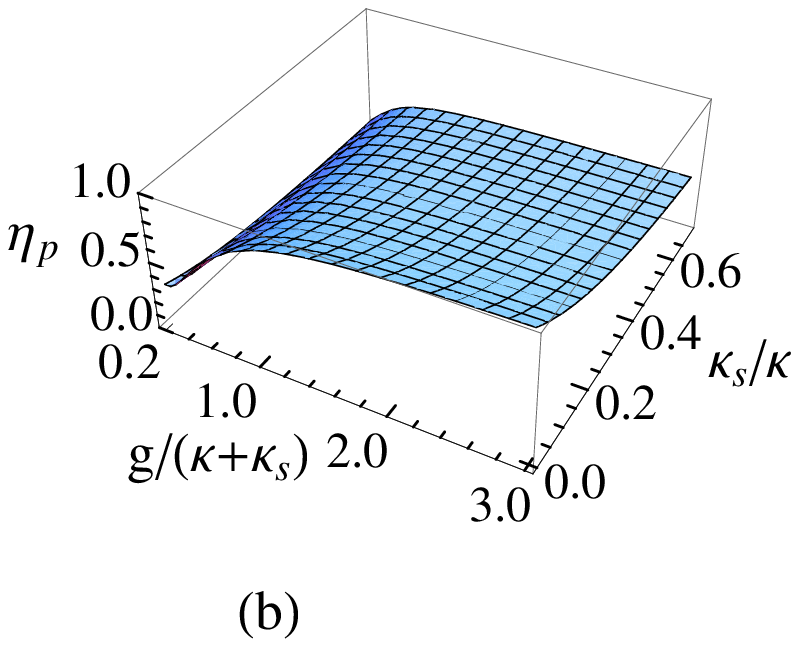}
\caption{ (Color online) Fidelity and efficiency of our P-S-QND (for the
even-parity mode) vs the coupling strength and cavity side leakage
rate with $\gamma=0.1\kappa$.} \label{figure7}
\end{figure}

\begin{figure}[htb]                    
\centering
\includegraphics[width=7 cm]{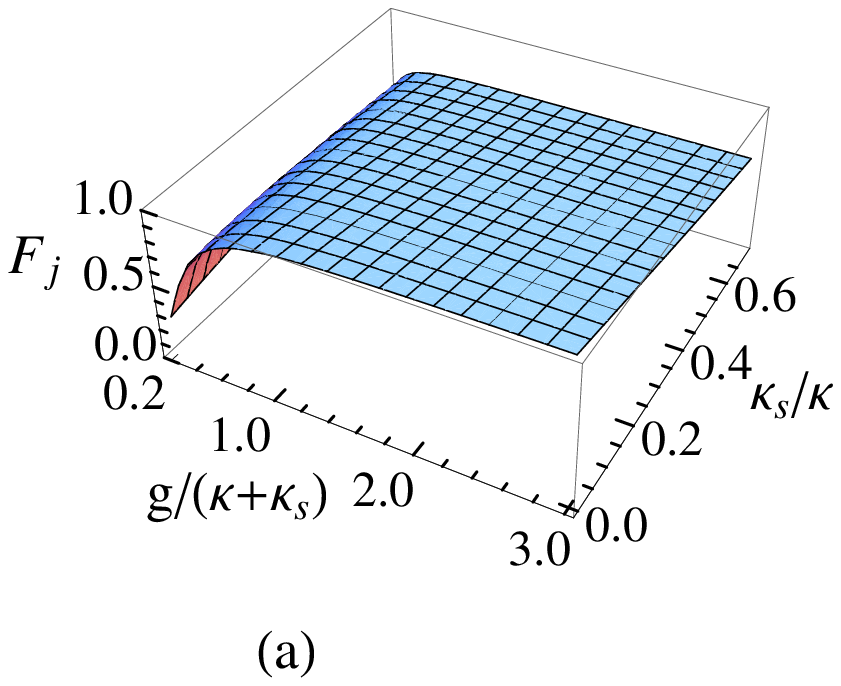}
\includegraphics[width=7 cm]{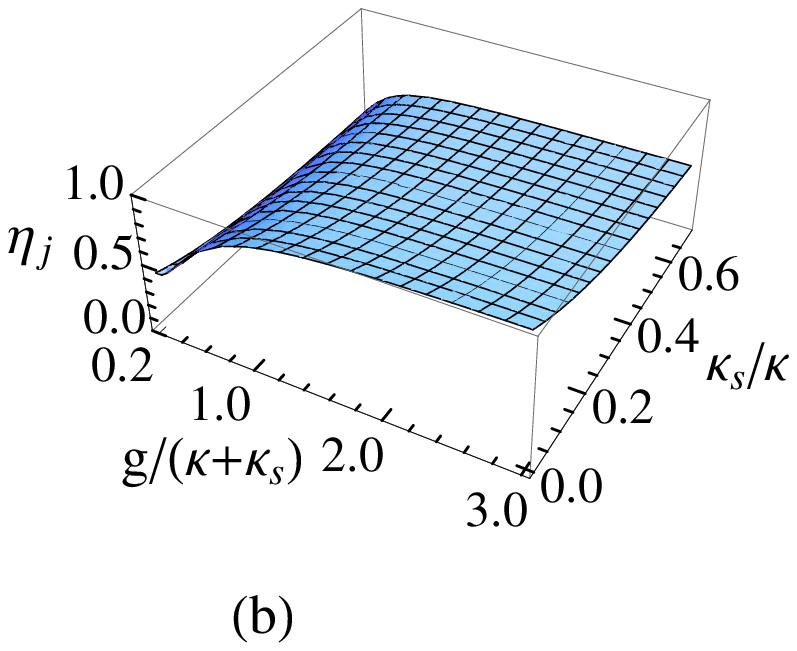}
\caption{ (Color online) Fidelity and efficiency of our QSJM (in the
present hyper-EPP) vs the coupling strength and cavity side leakage
rate with $\gamma=0.1\kappa$.} \label{figure8}
\end{figure}

In the resonant condition ($\omega_c=\omega_{X^-}=\omega_0=\omega$),
the fidelity of our proposal is mainly reduced by the cavity side
leakage and cavity coupling strength (discussed in Appendix
\ref{appendixb}). The fidelities and the efficiencies of the P-S-QND
and the QSJM are shown in Figs. \ref{figure7} and \ref{figure8}
with the coupling strength and cavity side leakage, respectively. Both of
these operations work efficiently with a strong coupling
strength and a low side leakage and cavity loss rate
$\kappa_s/\kappa$. The strong coupling strength
$g/(\kappa+\kappa_s)\simeq0.5$ has been observed in a $d=1.5$ $\mu$m
micropillar  microcavity \cite{couple} with a quality factor of
$Q\sim8800$. By improving the sample designs, growth, and
fabrication of QDs \cite{couple2}, the coupling strength
$g/(\kappa+\kappa_s)\simeq2.4$ ($Q\sim4\times10^4$) has been
achieved \cite{couple1} in a $d=1.5$ $\mu$m micropillar microcavity.
The fidelities and the efficiencies of of P-S-QND ($F_p$, $\eta_p$) and QSJM ($F_j$, $\eta_j$) are
$F_p=90.4\%$, $\eta_p=39.6\%$ and $F_j=76.5\%$, $\eta_j=50\%$ in the
case with $g/(\kappa+\kappa_s)\simeq0.5$ and $\kappa_s/\kappa\simeq0.3$.
They are $F_p=92.6\%$, $\eta_p=60.3\%$ and $F_j=84.4\%$,
$\eta_j=68.5\%$ for the case with coupling strength
$g/(\kappa+\kappa_s)\simeq2.4$ and $\kappa_s/\kappa\simeq0.3$, and they are
$F_p=100\%$, $\eta_p=96.6\%$ and $F_j=99.1\%$, $\eta_j=97.4\%$ for the case with
side leakage and cavity loss rate $\kappa_s/\kappa\simeq0$
[$g/(\kappa+\kappa_s)\simeq2.4$]. The fidelities and the
efficiencies of the two operations are mainly reduced by a weak
coupling strength and a high cavity side leakage. In experiment,
the quality factor of a micropillar microcavity is dominated by the
side leakage and cavity loss rate $\kappa_s/\kappa$. The side
leakage and cavity loss rate $\kappa_s/\kappa\simeq0.7$ has been
achieved in a $d=1.5$ $\mu$m micropillar with a quality factor of
$Q\sim1.7\times10^4$ [$g/(\kappa+\kappa_s)\simeq1$] by thinning down
the top mirrors \cite{QD5}. In this case, the fidelities and the
efficiencies of the two operations are $F_p=78.7\%$,
$\eta_p=39.2\%$ and $F_j=65.9\%$, $\eta_j=49.6\%$.

With the P-S-QND and the QSJM, we construct a two-step hyper-EPP for
the mixed hyperentangled Bell states with polarization bit-flip
errors and spatial-mode phase-flip errors. In the first step, Alice and
Bob both perform the P-S-QNDs on their photon pairs.
The fidelity and efficiency of this step are $F_1=92.7\%$ and
$\eta_1=47.6\%$ in the case with $g/(\kappa+\kappa_s)\simeq2.4$ and $\kappa_s/\kappa\simeq0.2$,
and they are $F_1=98\%$ and $\eta_2=65\%$ in the case with $g/(\kappa+\kappa_s)\simeq2.4$
and $\kappa_s/\kappa\simeq0.1$ (with P-S-QNDs in the even-parity mode).
In the second step, Alice and Bob both perform the
QSJM on their photon pairs. The fidelity and
efficiency of this step are $F_2=79.8\%$ and $\eta_2=57.3\%$ in the case with
$g/(\kappa+\kappa_s)\simeq2.4$ and $\kappa_s/\kappa\simeq0.2$, and they are
$F_2=88.9\%$ and $\eta_2=72.3\%$ in the case with $g/(\kappa+\kappa_s)\simeq2.4$
and $\kappa_s/\kappa\simeq0.1$. [Photons $C$ and $D$ in cases (3) and
(4) are detected in the first step to increase the fidelity and the efficiency
of the second step, which is identical to detecting photons $C$ and $D$ in the second step.]
The efficiency of the hyper-EPP is largely reduced by the cavity side leakage.
In order to iterate the hyper-EPP  process for obtaining high-fidelity
hyperentangled Bell states, the cavity side leakage should be small in the
strong-coupling regime, which means high-efficiency operations in experiment
are required to get a stronger coupling strength with a lower side leakage
in micropillars.

In the hyper-EPP without QSJM, the photonic states with only one DOF
in the preserving condition have to be discarded. In the present
hyper-EPP, the photonic entangled state with only polarization
(spatial-mode) DOF in the preserving condition can transfer its
polarization (spatial-mode) state to another photonic state with the
spatial-mode (polarization) DOF in the preserving condition, which
can preserve more high-fidelity entangled photon pairs. That is, the
efficiency of a hyper-EPP can be improved by using our
QSJM. Moreover, we generalize our hyper-EPP for mixed hyperentangled
GHZ states with polarization bit-flip errors and spatial-mode
phase-flip errors, which is more complicated than the one for mixed
hyperentangled Bell states. This generalization shows that our hyper-EPP with QSJM
can also be used to improve the efficiency of the hyper-EPP for mixed
multiphoton hyperentangled states.

In summary, we have investigated the possibility of improving the
efficiency of the hyper-EPP for mixed hyperentangled states with
polarization bit-flip errors and spatial-mode phase-flip errors,
resorting to the P-S-QND and the QSJM that are constructed with the
transmission-reflection rule of double-sided QD-cavity systems. The
present two-step hyper-EPP can greatly improve the efficiency by
preserving the states that are discarded in the hyper-EPP without
QSJM. We have analyzed the experimental feasibility of the two steps
of our hyper-EPP, and the analysis shows that they work efficiently in the
strong-coupling regime with low cavity side leakage. The present
two-step hyper-EPP with our QSJM can be generalized to improve the
efficiency of the hyper-EPP for mixed multiphoton hyperentangled
states, and it is useful for improving the entanglement of photon
systems with several DOFs in long-distance high-capacity quantum
communication.

\section*{ACKNOWLEDGMENTS}

This work is supported by the National Natural Science Foundation of
China under Grants No. 11174039 and No. 11474026 and NECT-11-0031.

\appendix

\section{QSJM for hyperentangled Bell states}
\label{appendixa}

In this appendix, we discuss the QSJM for hyperentangled Bell states
in detail. Let us  suppose that the two photon pairs, $AB$ and $A'B'$,
are in different polarization parity modes [Photons
$C$, $D$, $C'$, $D'$ in cases (3) and (4) are detected in the first step,
which is identical to detecting photons $C$, $D$, $C'$, $D'$ in the second
step]. That is,
\begin{eqnarray}                           
|\phi\rangle_{AB}&=&\frac{1}{2}(|RL\rangle+|LR\rangle)_{AB}(|a_1b_1\rangle+|a_2b_2\rangle),\nonumber\\
|\phi\rangle_{A'B'}&=&\frac{1}{2}(|RR\rangle+|LL\rangle)_{A'B'}(|a'_1b'_2\rangle+|a'_2b'_1\rangle).\nonumber\\
\end{eqnarray}

Alice first performs the Hadamard operations on the polarization DOF
of photon $A'$, and then she puts photon $A'$ into the
quantum circuit shown in Fig. \ref{figure2}(a). After these
operations are performed on photon $A'$, Bob performs the same
operations as those performed by Alice on photon $B'$. After
photon $B'$ passes through the quantum circuit, Alice and Bob perform the Hadamard operations on electron spins $e_a$ and $e_b$,
and the state of the quantum system composed of QDs and photons
$A'$ and $B'$ is transformed from $|\Phi_0\rangle_{A'B'e_ae_b}$ to
$|\Phi_1\rangle_{A'B'e_ae_b}$. Here
\begin{eqnarray}                           
|\Phi_0\rangle_{A'B'e_ae_b}&=&\frac{1}{2}(|\uparrow\rangle+|\downarrow\rangle)_{e_a}\otimes(|\uparrow\rangle+|\downarrow\rangle)_{e_b}\nonumber\\
&&\otimes|\phi\rangle_{A'B'},\nonumber\\\
|\Phi_1\rangle_{A'B'e_ae_b}&=&\frac{1}{4}[(|RR\rangle+|LL\rangle)_{A'B'}(|\uparrow\uparrow\rangle+|\downarrow\downarrow\rangle)_{e_ae_b}\nonumber\\
&&+(|RL\rangle+|LR\rangle)_{A'B'}(|\uparrow\uparrow\rangle-|\downarrow\downarrow\rangle)_{e_ae_b}]\nonumber\\
&&\otimes(|a'_1b'_2\rangle+|a'_2b'_1\rangle).
\end{eqnarray}
The polarization states of photons $A'$ and $B'$ are measured in the
orthogonal basis $\{|R\rangle,|L\rangle\}$. If the two photons are
in the even-parity polarization mode, the polarization state of the
two-photon system $A'B'$ is transferred to the state of electron
spins $e_a$ and $e_b$. If the polarization state of the photon pair
$A'B'$ is in the odd-parity polarization mode, a phase-flip
operation $\sigma_z=|\uparrow\rangle\langle\uparrow| -
|\downarrow\rangle\langle\downarrow|$ must be performed on electron
spin $e_a$.

Subsequently, photon $A$ is put into the quantum circuit shown
in Fig. \ref{figure2}(a) by Alice. After photon $A$ passes
through the quantum circuit, Alice performs the Hadamard operations
on the polarization DOF of photon $A$ and electron spin
$e_a$. Then photon $A$ is put into the quantum circuit again.
After these operations  are performed on photon $A$, Bob
performs the same operations as those performed by Alice on photon $B$ and electron spin
$e_b$. Finally, Alice and Bob perform
the Hadamard operations on electron spins $e_a$ and $e_b$, and the
state of the quantum system composed of QDs and photons $A$ and $B$ are
transformed from $|\Phi_1\rangle_{ABe_ae_b}$ to
$|\Phi_2\rangle_{ABe_ae_b}$. Here
\begin{eqnarray}                           
|\Phi_1\rangle_{ABe_ae_b}&=&\frac{1}{\sqrt{2}}(|\uparrow\uparrow\rangle+|\downarrow\downarrow\rangle)_{e_ae_b}\otimes|\phi\rangle_{AB},\nonumber\\\
|\Phi_2\rangle_{ABe_ae_b}&=&\frac{1}{2\sqrt{2}}(|RR\rangle-|LL\rangle)_{AB}(|\uparrow\downarrow\rangle+|\downarrow\uparrow\rangle)_{e_ae_b}\nonumber\\
&&\otimes(|a_1b_1\rangle+|a_2b_2\rangle).
\end{eqnarray}
Electron spins $e_a$ and $e_b$ are measured in the orthogonal basis
$\{|\uparrow\rangle,|\downarrow\rangle\}$. If the two electron spins
are in the odd-parity mode, a phase-flip operation $\sigma^P_z$
performed on photon $A$ is required to transfer the state of
electron spins $e_a$ and $e_b$ to the polarization state of photon
pair $AB$.

If the two photon pairs $AB$ and $A'B'$
are in the same polarization parity mode,
\begin{eqnarray}                           
|\phi'\rangle_{AB}&=&\frac{1}{2}(|RR\rangle+|LL\rangle)_{AB}(|a_1b_1\rangle+|a_2b_2\rangle),\nonumber\\
|\phi'\rangle_{A'B'}&=&\frac{1}{2}(|RR\rangle+|LL\rangle)_{A'B'}(|a'_1b'_2\rangle+|a'_2b'_1\rangle),\nonumber\\
\end{eqnarray}
Alice and Bob can perform the same QSJM operation on the two photon
pairs. The finial state $|\Phi'_2\rangle_{ABe_ae_b}$ will become
\begin{eqnarray}                           
|\Phi'_2\rangle_{ABe_ae_b}&=&\frac{1}{2\sqrt{2}}(|RR\rangle+|LL\rangle)_{AB}(|\uparrow\uparrow\rangle+|\downarrow\downarrow\rangle)_{e_ae_b}\nonumber\\
&&\otimes(|a_1b_1\rangle+|a_2b_2\rangle).
\end{eqnarray}
The two electron spins
are in the even-parity mode, and the polarization state of the photon pair $A'B'$
is transferred  to the polarization state of the photon
pair $AB$.

If the two photon pairs $AB$ and $A'B'$
are in the different polarization parity modes,
\begin{eqnarray}                           
|\phi'\rangle_{AB}&=&\frac{1}{2}(|RR\rangle+|LL\rangle)_{AB}(|a_1b_1\rangle+|a_2b_2\rangle),\nonumber\\
|\phi''\rangle_{A'B'}&=&\frac{1}{2}(|RL\rangle+|LR\rangle)_{A'B'}(|a'_1b'_2\rangle+|a'_2b'_1\rangle),\nonumber\\
\end{eqnarray}
Alice and Bob can also perform the same QSJM operation on the two
photon pairs. The finial state $|\Phi''_2\rangle_{ABe_ae_b}$ will
become
\begin{eqnarray}                           
|\Phi''_2\rangle_{ABe_ae_b}&=&\frac{1}{2\sqrt{2}}(|RR\rangle-|LL\rangle)_{AB}(|\uparrow\uparrow\rangle+|\downarrow\downarrow\rangle)_{e_ae_b}\nonumber\\
&&\otimes(|a_1b_1\rangle+|a_2b_2\rangle).
\end{eqnarray}
The two electron spins
are in the even-parity mode, and the polarization state of the photon pair $A'B'$
is transferred to the polarization state of the photon
pair $AB$ after the Hadamard operations are performed on the polarization
DOF of the photon pair $AB$.

If both photon pairs $AB$ and $A'B'$
are in the same polarization parity mode,
\begin{eqnarray}                           
|\phi''\rangle_{AB}&=&\frac{1}{2}(|RL\rangle+|LR\rangle)_{AB}(|a_1b_1\rangle+|a_2b_2\rangle),\nonumber\\
|\phi'''\rangle_{A'B'}&=&\frac{1}{2}(|RL\rangle+|LR\rangle)_{A'B'}(|a'_1b'_2\rangle+|a'_2b'_1\rangle),\nonumber\\
\end{eqnarray}
Alice and Bob can perform the same QSJM operation on the two photon
pairs as well. The finial state $|\Phi'''_2\rangle_{ABe_ae_b}$ will
become
\begin{eqnarray}                           
|\Phi'''_2\rangle_{ABe_ae_b}&=&\frac{1}{2\sqrt{2}}(|RR\rangle+|LL\rangle)_{AB}(|\uparrow\downarrow\rangle+|\downarrow\uparrow\rangle)_{e_ae_b}\nonumber\\
&&\otimes(|a_1b_1\rangle+|a_2b_2\rangle).
\end{eqnarray}
 The two electron spins
are in the odd-parity mode, and the polarization state of photon
pair $A'B'$ is transferred to the polarization state of photon pair
$AB$ after a phase-flip operation $\sigma^P_z$ is performed on
photon $A$ and  the Hadamard operations are performed on the
polarization DOF of photon pair $AB$.

\section{Fidelities and efficiencies of P-S-QND and QSJM}
\label{appendixb}

In the resonant condition ($\omega_c=\omega_{X^-}=\omega_0=\omega$),
the fidelity of our proposal is mainly reduced by the cavity side
leakage and cavity coupling strength. The transmission and
reflection rule in Eq. (\ref{eq4}) is reduced to
\begin{eqnarray} 
|R^\uparrow, i_2, \uparrow\rangle      &\rightarrow&   |r||L^\downarrow, i_2, \uparrow\rangle   - |t||R^\uparrow, i_1, \uparrow\rangle,\nonumber\\
|L^\downarrow, i_1,  \uparrow\rangle   &\rightarrow&   |r||R^\uparrow, i_1, \uparrow\rangle     - |t||L^\downarrow, i_2,  \uparrow\rangle,\nonumber\\
|R^\downarrow, i_1, \downarrow\rangle  &\rightarrow&   |r||L^\uparrow, i_1, \downarrow\rangle   - |t||R^\downarrow, i_2, \downarrow\rangle,\nonumber\\
|L^\uparrow , i_2,  \downarrow\rangle  &\rightarrow&   |r||R^\downarrow, i_2, \downarrow\rangle - |t||L^\uparrow , i_1, \downarrow\rangle,\nonumber\\
|R^\downarrow, i_1, \uparrow\rangle    &\rightarrow&  -|t_0||R^\downarrow, i_2, \uparrow\rangle  + |r_0||L^\uparrow, i_1, \uparrow\rangle,\nonumber\\
|L^\uparrow , i_2,  \uparrow\rangle    &\rightarrow&  -|t_0||L^\uparrow , i_1,  \uparrow\rangle  + |r_0||R^\downarrow, i_2,  \uparrow\rangle,\nonumber\\
|R^\uparrow, i_2, \downarrow\rangle    &\rightarrow&  -|t_0||R^\uparrow, i_1, \downarrow\rangle  + |r_0||L^\downarrow, i_2, \downarrow\rangle,\nonumber\\
|L^\downarrow, i_1, \downarrow\rangle  &\rightarrow&
-|t_0||L^\downarrow, i_2, \downarrow\rangle+|r_0||R^\uparrow, i_1,
\downarrow\rangle.
\end{eqnarray}

The fidelity of a quantum information process is defined as
$F=|\langle\psi_f|\psi\rangle|^2$, where $|\psi\rangle$ is the ideal
final state of the system after the quantum information processing
and $|\psi_f\rangle$ is the final state of the system found by considering
an experimental environment. The efficiency of a photonic quantum
information process is defined as the probability of the photons to
being detected after quantum information processing in a
practical experimental environment. The fidelity and the efficiency
of our P-S-QND (for the even-parity mode) can be described as
\begin{eqnarray}                           
F_p&=&\frac{\left(\sum\limits_{i=1}^4 m_i\right)^2}{4\left(\sum\limits_{j=1}^4m_j^2+\sum\limits_{k=5}^9m_k\right)},\nonumber\\
\eta_p&=&\frac{1}{16}(|r|^2+|t|^2+|t_0|^2+|r_0|^2)^4.\;\;\;\;\;\;\;\;
\end{eqnarray}
The fidelity and the efficiency of our QSJM (in the present
hyper-EPP process) can be described as
\begin{eqnarray}                           
F_j&=&\frac{(|r|+|t|+|r_0|+|t_0|)^2(|r|+|t_0|)^4}{2\sum\limits_{k=1}^8n_k^2\left[(|r|+|t|)^2+(|r_0|+|t_0|)^2\right]},\nonumber\\
\eta_j&=&\frac{1}{8}(|r|^2+|t|^2+|t_0|^2+|r_0|^2)^3.\;\;\;\;\;\;\;\;\;\;\;\;\;\;\;\;
\end{eqnarray}
Here
\begin{eqnarray}                          
m_1&=&(|t|^2+|r|^2)|t_0|^2+(|t_0|^2+|r_0|^2)|r_0|^2,\nonumber\\
m_2&=&(|t|^2+|r|^2)|r|^2+(|t_0|^2+|r_0|^2)|t|^2, \nonumber\\
m_3&=&(|t|^2+|r|^2)|r_0|^2+(|t_0|^2+|r_0|^2)|t_0|^2,\nonumber\\
m_4&=&(|t|^2+|r|^2)|t|^2+(|t_0|^2+|r_0|^2)|r|^2,\nonumber\\
m_5&=&2(|t_0|^2|r_0|^2+|r|^2|t|^2)(|t_0|^2+|t|^2+|r_0|^2+|r|^2)^2,\nonumber\\
m_6&=&4|t_0|^2|r|^2(|t|^2+|r_0|^2)^2,\nonumber\\
m_7&=&4|t_0|^2|t|^2(|r_0|^2+|r|^2)^2,\nonumber\\
m_8&=&4|r_0|^2|r|^2(|t_0|^2+|t|^2)^2,\nonumber\\
m_9&=&4|r_0|^2|t|^2(|t_0|^2+|r|^2)^2,\nonumber\\
n_1&=&(|t|+|r|)|r|+(|r|-|t|)|t|+(|t_0|+|r_0|)|r|\nonumber\\
&&-(|t_0|-|r_0|)|t|,\nonumber\\
n_2&=&(|t|+|r|)|t|+(|r|-|t|)|r|+(|t_0|+|r_0|)|t|\nonumber\\
&&-(|t_0|-|r_0|)|r|,\nonumber\\
n_3&=&(|t_0|+|r_0|)|t_0|-(|r_0|-|t_0|)|r_0|+|(|t|+|r|)|t_0|\nonumber\\
&&-(|r|-|t|)|r_0|,\nonumber\\
n_4&=&(|t_0|+|r_0|)|r_0|-(|r_0|-|t_0|)|t_0|+|(|t|+|r|)|r_0|\nonumber\\
&&-(|r|-|t|)|t_0|,\nonumber\\
n_5&=&(|t|+|r|)|t_0|+(|r|-|t|)|r_0|+(|t_0|+|r_0|)|t_0|\nonumber\\
&&-(|t_0|+|r_0|)|r_0|,\nonumber\\
n_6&=&(|t|+|r|)|r_0|+(|r|-|t|)|t_0|+(|t_0|+|r_0|)|r_0|\nonumber\\
&&-(|t_0|+|r_0|)|t_0|,\nonumber\\
n_7&=&(|t_0|+|r_0|)|r|-(|r_0|-|t_0|)|t|+|(|t|+|r|)|r|\nonumber\\
&&-(|r|-|t|)|t|,\nonumber\\
n_8&=&(|t_0|+|r_0|)|t|-(|r_0|-|t_0|)|r|+|(|t|+|r|)|t|\nonumber\\
&&-(|r|-|t|)|r|.
\end{eqnarray}

\end{document}